\documentclass[%
 aip,
 amsmath,amssymb,
 reprint,%
]{revtex4-1}

\usepackage{graphicx}
\usepackage{dcolumn}
\usepackage{bm}

\usepackage[utf8]{inputenc}
\usepackage[T1]{fontenc}
\usepackage{mathptmx}
\usepackage{etoolbox}
\usepackage{bigints}
\usepackage[dvipsnames]{xcolor}
\usepackage{comment} 

\makeatletter
\def\@email#1#2{%
 \endgroup
 \patchcmd{\titleblock@produce}
  {\frontmatter@RRAPformat}
  {\frontmatter@RRAPformat{\produce@RRAP{*#1\href{mailto:#2}{#2}}}\frontmatter@RRAPformat}
  {}{}
}%
\makeatother
\begin{document}

\preprint{AIP/123-QED}

\title[Local bond order parameters for hydrates]{Rotationally invariant local bond order parameters for accurate determination of hydrate structures}


\author{Iván M. Zerón}
\affiliation{Laboratorio de Simulaci\'on Molecular y Qu\'imica Computacional, CIQSO-Centro de Investigaci\'on en Qu\'imica Sostenible and Departamento de Ciencias Integradas, Universidad de Huelva, 21006 Huelva Spain}

\author{Jesús Algaba}
\affiliation{Laboratorio de Simulaci\'on Molecular y Qu\'imica Computacional, CIQSO-Centro de Investigaci\'on en Qu\'imica Sostenible and Departamento de Ciencias Integradas, Universidad de Huelva, 21006 Huelva Spain}

\author{José Manuel M\'{\i}guez}
\affiliation{Laboratorio de Simulaci\'on Molecular y Qu\'imica Computacional, CIQSO-Centro de Investigaci\'on en Qu\'imica Sostenible and Departamento de Ciencias Integradas, Universidad de Huelva, 21006 Huelva Spain}

\author{Bruno Mendiboure}
\affiliation{Laboratoire des Fluides Complexes et Leurs R\'eservoirs, UMR5150, Universit\'e de Pau et des Pays de l’Adour, B.P. 1155, Pau Cdex 64014, France}

\author{Felipe J. Blas$^{*}$}
\affiliation{Laboratorio de Simulaci\'on Molecular y Qu\'imica Computacional, CIQSO-Centro de Investigaci\'on en Qu\'imica Sostenible and Departamento de Ciencias Integradas, Universidad de Huelva, 21006 Huelva Spain}

\begin{abstract}
\textbf{ABSTRACT}

\noindent
Averaged local bond order parameters based on spherical harmonics, also known as Lechner and Dellago order parameters, are routinely used to determine crystal structures in molecular simulations. Among different options, the combination of the $\overline{q}_{4}$ and $\overline{q}_{6}$ parameters is one of the best choices in the literature since allows one to distinguish, not only between solid- and liquid-like particles but also between different crystallographic phases, including cubic and hexagonal phases. Recently, Algaba \emph{et al.} [\emph{J. Colloid Interface Sci.}~\textbf{623}, 354, (2022)] have used the Lechner and Dellago order parameters to distinguish hydrate- and liquid-like water molecules in the context of determining the carbon dioxide hydrate-water interfacial free energy. According to the results, the preferred combination previously mentioned is not the best option to differentiate between hydrate- and liquid-like water molecules. In this work, we revisit and extend the use of these parameters to deal with systems in which clathrate hydrates phases coexist with liquid phases of water. We consider carbon dioxide, methane, tetrahydrofuran, nitrogen, and hydrogen hydrates that exhibit sI and sII crystallographic structures. We find that the $\overline{q}_{3}$ and $\overline{q}_{12}$ combination is the best option possible between a large number of possible different pairs to distinguish between hydrate- and liquid-like water molecules in all cases. 

\end{abstract}


\maketitle
$^*$Corresponding author: felipe@uhu.es

%

\section{Introduction}

When a solid is heated, at constant pressure, it melts and transforms into a liquid via a first-order transition. This is observed from both experiments and computer simulations.~\cite{Debenedetti1996a} However, the formation of a solid from a bulk liquid phase is not straightforward. In other words, a solid-fluid first-order transition is not symmetric in both directions.~\cite{Debenedetti1996a} Whereas it is difficult to maintain stable a solid at temperatures above the melting temperature, it is possible to observe a metastable liquid phase at temperatures well below the melting point of the system although the solid is the thermodynamically stable phase. For instance, in a series of beautiful experiments, Kanno \emph{et al.}~\cite{Kanno1975a} were able to stabilize liquid water up to $-32^{\circ}\,\text{C}$ at ambient pressure and $-92^{\circ}\,\text{C}$ at $2000\,\text{bar}$. Why do substances generally not crystallize at temperatures below the melting temperature and stay in the metastable liquid phase? The existence of a metastable phase at conditions for which the thermodynamic stable phase is a solid generally occurs under homogeneous conditions, i.e., in the absence of impurities or surfaces in the system. In this case, the formation of a solid occurs via homogeneous nucleation.~\cite{Debenedetti1996a,Kashchiev2000a} The name is used to differentiate from the formation of the solid phase due to the presence of impurities or solid surfaces. In this case, the formation of the solid is denoted as heterogeneous nucleation.~\cite{Kashchiev2000a} According to the well-accepted vision of a solid formation, this process is a two-well-differentiated step: homogeneous nucleation and growth mechanism.~\cite{Grabowska2022a} 

Homogeneous nucleation is an activated process in which the system must overcome a free energy barrier. The existence of this energy barrier is a consequence of the microscopic mechanism of solid formation. It is well established that it is necessary to initially form an embryo, with a spherical or pseudo-spherical shape, of the new stable phase (solid) inside the metastable phase (liquid), for the appearance of a solid phase.~\cite{Debenedetti1996a,Kashchiev2000a} This occurs spontaneously via fluctuations in the system. However, the formation of this nucleus has an energetic penalty: the energy cost of the formation of a solid-liquid interface. Thus, homogeneous nucleation can be viewed as an energetic competition between two contributions to the free energy: a negative (decreasing) contribution due to the transfer of $N$ molecules from the liquid to the solid state, $\sim N\Delta\mu$, and a positive (increasing) contribution due to cost in the creation of surface area $\mathcal{A}$ of the spherical or pseudo-spherical nucleus, $\sim\gamma\mathcal{A}$.~\cite{Auer2001a} Here $\Delta\mu=\mu_{\text{S}}-\mu_{\text{L}}$ is the difference in chemical potential between the solid (stable) and the liquid (metastable) states, also known as the driving force for nucleation, and $\gamma$ is the interfacial free energy of the solid-liquid interface. When this nucleus of the emerging stable phase is larger than a certain critical size (critical nucleus), the solid phase grows spontaneously. 

According to Classical Nucleation Theory (CNT),~\cite{Volmer1926a,Becker1935a,Kelton1991a} there exist three key magnitudes in homogeneous nucleation: the driving force for nucleation, $\Delta\mu$, the solid-liquid interfacial free energy, $\gamma$, and the nucleation rate, $J$. These three properties can be determined from experimental data and computer simulation.~\cite{Kashchiev2000a} However, not all of them are easy to evaluate from both methods and the accuracy of the predictions depends on the property and the thermodynamic condition at which are calculated. In this work, we concentrate on some technical calculations needed to estimate nucleation rates and interfacial free energies from computer simulation. Whereas $\Delta\mu$ is relatively easy to evaluate from computer simulation, the estimation of $J$ and $\gamma$ from simulation is a challenging problem.

The nucleation rate, one of critical magnitude in nucleation studies, is formally defined as the number of critical clusters formed per unit of volume and per unit of time. Unfortunately, homogeneous nucleation is a rare event.~\cite{Debenedetti1996a} This implies that nucleation takes a very long time to occur spontaneously, particularly in simulations. It is possible to measure nucleation rates from experiments.~\cite{Debenedetti1996a,Kashchiev2000a} Typically, nucleation rates found in experiments are in the range $\sim 10^{0-14}/(\text{m}^{3}\,\text{s})$. This corresponds to observation times of minutes and sample sizes between $10^{-3}$ and $10^{-17}\,\text{m}^{3}$.~\cite{Debenedetti1996a,Grabowska2023a} However, simulation time scales are very different due to limitations of the system size and the available simulation time. For instance, in brute-force (BF) simulations the nucleation rate is of the order of $10^{30-33}/(\text{m}^{3}\,\text{s})$.~\cite{Espinosa2014b,Espinosa2016d}

There are different techniques in simulation to estimate nucleation rates. It is possible to observe spontaneous nucleation at low temperatures by using BF simulations. Unfortunately, these simulations cannot provide estimates of nucleation rates found in experiments because they need extremely high driving forces and consequently, very low temperatures. To overcome this, several enhanced sampling techniques have been introduced to specifically deal with rare-event problems, such as Forward Flux Sampling,~\cite{Allen2005a} Transition Path Sampling,~\cite{Bolhuis2002a,Lechner2011a,Beckham2011a} Metadynamics,~\cite{Laio2002a,Trudu2006a} Lattice Mold,~\cite{Espinosa2016e} and Umbrella Sampling.~\cite{Torrie1977a,Auer2001a} Although these techniques are much faster than brute-force simulations, they still require substantial computational resources. As a consequence, most nucleation studies focus on only a few selected state points. In recent years, an alternative methodology has been improved and developed to estimate nucleation rates at higher temperatures: the Seeding technique~\cite{Espinosa2016c,Bai2005a,Bai2006a,Knott2012a,Sanz2013a} in combination with CNT.~\cite{Volmer1926a,Becker1935a,Kelton1991a} Different authors have used this technique to determine nucleation rates of water, NaCl, and hydrates.~\cite{Jacobson2011a,Knott2012a,Sanz2013a,Espinosa2014b,Bianco2021a,Lamas2021a,Grabowska2023a}

Most of these techniques need an order parameter to distinguish between the nucleus (stable phase) and the surrounding metastable phase. Different order parameter choices yield different nucleus sizes and, consequently, different nucleation rates. The sensitivity of $J$ is different depending on the method.~\cite{Gispen2024a} But, in general, the final estimation of the nucleation rate largely depends on the order parameter choice. It is important to recall here that some of the special techniques described above require a ``reaction coordinate'' that measures the degree of crystallinity of the system as it moves from the liquid to the solid phase to estimate nucleation barriers and compute the crossing rate. In most cases, this reaction coordinate is identified as an appropriate order parameter.~\cite{tenWolde1995a,tenWolde1996a}

The solid-liquid interfacial free energy, another critical magnitude in nucleation studies, is defined as the work needed to form a thin crystalline slab of solid phase in the liquid bulk phase per interfacial area. There exist different computer simulation techniques, including the Cleaving,~\cite{Broughton1986a,Davidchack2000a,Davidchack2003a,Davidchack2010a,Davidchack2012a} Capillary Fluctuation,~\cite{Hoyt2001a,Benet2014a} Metadynamics~\cite{Angioletti-Uberti2010a} or Tethered Monte Carlo methods.~\cite{Fernandez2012a} One of the most recent methodologies proposed in the literature for addressing solid-fluid interfacial free energies in both simple and complex systems, such as hard spheres,~\cite{Espinosa2014a} Lennard-Jones,~\cite{Espinosa2014a} water,\cite{Espinosa2016a} NaCl,~\cite{Soria2018a,Sanchez-Burgos2023a} and hydrates,\cite{Algaba2022b,Zeron2022a,Romero-Guzman2023a} is the Mold Integration technique.~\cite{Espinosa2014a} The method relies on using a mold with attractive interactive sites to induce the formation of a solid crystalline slab within the bulk liquid phase.~\cite{Espinosa2014a} One important issue is tracking the number of molecules in the solid crystalline slab when the mold is turned on. To determine this magnitude, an order parameter must be defined, as it happens in the case of nucleation, to differentiate between solid-like and liquid-like molecules.

According to the previous discussion, the common point of the simulation methods described above for estimating nucleation rates and interfacial free energies is the need to use order parameters to distinguish between solid-like and liquid-like molecules. Numerous order parameters are described in the literature to differentiate molecules in solid and liquid phases. The first set of order parameters for distinguishing between solid- and liquid-like particles was proposed by Steinhardt \emph{et al.}~\cite{Steinhardt1983a} based on the idea of Frank on local orientational symmetries of condensed phases.~\cite{Frank1952a} According to this, local orientational symmetries are important to characterize the internal structure of three-dimensional liquids and solids.  Steinhardt and coworkers~\cite{Steinhardt1983a} proposed to associate a set of spherical harmonics, $Y_{lm}(\mathbf{r}_{ij})$, with neighbors of a given atom defining the complex vector associated with a particle $i$,

\begin{equation}
q_{lm}(i)=\dfrac{1}{N_{b}(i)}\sum_{j=1}^{N_{b}(i)}Y_{lm}(\mathbf{r}_{ij})
\end{equation}

\noindent
Here, $N_{b}(i)$ is the number of nearest neighbors of particle $i$, $l$ is a free integer parameter, and $m$ is an integer that runs from $m=-l$ to $m=+l$. In order to efficiently distinguish, not only between solid- and liquid-like particles but only to differentiate between different crystallographic structures, Steinhardt \emph{et al.}~\cite{Steinhardt1983a} defined the so-called local bond order parameters,

\begin{equation}
q_l(i)=\sqrt{\dfrac{4\pi}{2l+1}\sum_{m=-l}^{l}|q_{lm}(i)|^2}
\label{order-parameters}
\end{equation}

\noindent
Depending on the $l$ value, these parameters allow to differentiate different crystal symmetries. Frenkel and collaborators~\cite{vanDuijneveldt1992a,tenWolde1995a,tenWolde1996a,Auer2001a,Auer2004a} have used the ideas of Steinhardt \emph{et al.}~\cite{Steinhardt1983a} to investigate homogeneous nucleation of several systems, finding that $q_{4}$ and $q_{6}$ values are good choices to distinguish between different solid structures. Desgranges and Delhommelle~\cite{Desgranges2008a} have improved the method of the local order parameters using a $q_{4}$--$q_{6}$ planar representation that allows to identification easily solid-and liquid-like particles.

Unfortunately, the local order parameters of Steinhardt \emph{et al.}~\cite{Steinhardt1983a} do not allow a clear distinction between different local crystalline structures due to the presence of thermal fluctuations that smear out the order parameter distributions. To avoid this effect, Lechner and Dellargo~\cite{Lechner2008a} proposed a new version of the order parameters of Steinhardt and collaborators~\cite{Steinhardt1983a} by averaging over the bond order parameters of nearest neighbor particles. This allows a substantial increase in the accuracy of determining the crystal structure of the system. From then, several authors have used the averaged local order parameters of Lechner and Dellago~\cite{Lechner2008a} and the $\overline{q}_{4}$--$\overline{q}_{6}$ planar representation to successfully distinguish between solid- and liquid-like molecules in nucleation problems and determining solid-fluid interfacial free energies. However, Algaba \emph{et al.}~\cite{Algaba2022b} have recently discovered that the successful $\overline{q}_{4}$--$\overline{q}_{6}$ planar representation is not the best combination of differentiate between hydrate-like and liquid-like water molecules. These authors have determined for the first time the CO$_{2}$ hydrate--water interfacial free energy using an extension of the Mold Integration technique of Espinosa \emph{et al.},~\cite{Espinosa2014a} the so-called Mold Integration-Host (MI-H).~\cite{Algaba2022b,Romero-Guzman2023a} According to the results obtained by them, also corroborated by Grabowska and collaborators~\cite{Grabowska2023a}, other combinations of the averages local bond order parameters provide a better distinction between hydrate- and liquid-like water molecules.

Clathrate hydrates are non-stochiometric crystalline inclusion compounds consisting of a network of hydrogen-bonded molecules (host) conforming cages in which small molecules (guest) can be encapsulated under the appropriate thermodynamic conditions. When the host molecule is water, these compounds are simply called hydrates.~\cite{Sloan2008a,Ripmeester2022a} The structure of the hydrates depends on the thermodynamic conditions, but moreover, depends on the guest molecule encapsulated inside the hydrate. The guest has a high impact on the stability of the hydrate. Hydrates of small molecules, such as methane (CH$_4$) or carbon dioxide (CO$_2$), often crystallize forming the so-called structure sI. Hydrates of medium molecules (i.e., butane, cyclopentane, or tetrahydrofuran) crystallize in structure sII,~\cite{Sloan2008a,Ripmeester2022a} although this structure is also stabilized by very small guests such as nitrogen (N$_2$) and hydrogen (H$_2$).~\cite{Barnes2013a} Finally,  when there is a mixture of large and small molecules the sH structure is the most common.~\cite{Sloan2008a,Ripmeester2022a}

In this work, we focus on the study and use of the well-established averaged bond local order parameters proposed by Lechner and Dellago.~\cite{Lechner2008a} However, it is also possible to use more sophisticated methods for the classification of different local structures, including the K-nearest neighbors method.~\cite{Eppstein1997a,Zou2024a} It is also worth mentioning that the CHILL and CHILL+ algorithms of Molinero and collaborators are also valuable techniques for identifying crystallographic structures.~\cite{Moore2010b,Moore2011c,Nguyen2015a} In fact, one of the advantages of CHILL algorithms over the Lechner and Dellago order parameters is their ability to identify the cage formation of crystalline solid structures, including Ih and Ic ices, as well as clathrate hydrates. It is important to take into account that CHILL algorithms use the original (non-averaged) $q_{3}$ and $q_{4}$ Steinhardt order parameters to this end.~\cite{Moore2010b} According to Lechner and Dellago~\cite{Lechner2008a} and our previous and current works,~\cite{Algaba2022b,Zeron2022a,Romero-Guzman2023a,Grabowska2023a,Torrejon2024a}  the averaged order parameters of superior to distinguish between solid-like and liquid-like water molecules when dealing with hydrates. However, it is also interesting to use the CHILL algorithms to identify cages of hydrates. Unfortunately, the use of these two alternative strategies is out of the scope of this work.

The main goal of this work is to revisit and extend the use of the Lechner and Dellago~\cite{Lechner2008a} averaged local bond order parameters to deal with systems in which clathrate hydrate phases coexist with liquid phases of water. Particularly, we consider using these local order parameters in the context of estimating homogeneous nucleation rates and determining hydrate-water interfacial free energy using the Mold Integration technique. Note however that the use of averaged local bond order parameters could be useful in a great variety of methodologies, including rare-event techniques, Seeding, Lattice Mold, and Mold Integration methods, among many others.

The organization of this paper is as follows: In Sec. II, we describe the methodology used in the manuscript. Simulation details and molecular models are described in Sec. III. The results obtained in this work are discussed in detail in Sec. IV. Finally, conclusions are presented in Sec. V.

\section{Methodology}


In order to study the crystallization/melting process of a liquid/crystal phase, it is necessary to distinguish between liquid-like and crystal-like molecules.  Lechner and Dellago\cite{Lechner2008a} proposed a modification of the local bond order parameters proposed by Steinhardt \emph{et al.}~\cite{Steinhardt1983a} to distinguish between crystal-like and liquid-like molecules by the analysis of the local environment of each molecule of the system. The analysis of the environment of each molecule is based on the determination of the spherical harmonics which depend on the
angles between the vectors from the chosen molecule to its neighbors.  The local bond order parameters proposed by Steinhardt \emph{et al.}~\cite{Steinhardt1983a} are calculated as given by Eq.~\eqref{order-parameters}.

%

\noindent As we have already mentioned, $q_{lm}(i)$ is a complex vector calculated as a function of the spherical harmonics between the molecule $i$ and its neighbors, and $l$ and $m$ are the degree and order of the spherical harmonic functions associated with $q_{lm}(i)$ (we refer the lector to the original work of Steinhardt \emph{et al.}~\cite{Steinhardt1983a} for further information).

There are two main advantages when Steinhardt parameters are used to distinguish between crystal-like and liquid-like molecules. The first one is that these parameters are independent of a specific crystal structure and the second one is that we can tune the sensibility of these parameters by choosing different values of $l$, allowing us to label molecules not only as liquid or crystal but also to distinguish between different crystal structures. However, the main disadvantage of Steinhardt local bond order parameters is the lack of statistics due to the low number of neighbor molecules. As a result, thermal fluctuations can distort the order parameter distribution, or in other words, thermal fluctuations can cause that the parameters calculated from molecules in the liquid and crystal phases to become indistinguishable. 

Lechner and Dellago\cite{Lechner2008a} proposed a modification of the Steinhardt local bond order parameters to improve the accuracy of these parameters. Once the complex vector $q_{lm}(i)$ is calculated for molecule $i$ and all its neighbors, a new complex vector $\overline{q}_{lm}(i)$ is calculated by averaging the complex vectors, $q_{lm}(i)$, calculated for molecule $i$ and its neighbors. This is a simple and efficient method to improve the accuracy of the Steinhardt parameters and requires a minimum extra effort since the $q_{lm}(i)$ complex vectors of all the molecules of the system have already been calculated to apply the Steinhardt parameters. Once $\overline{q}_{lm}(i)$ is calculated, it is used directly on Eq.~\eqref{order-parameters} instead of $q_{lm}(i)$, obtaining the final expression proposed by Lechner and Dellago\cite{Lechner2008a} (we refer the lector to the original work for further information): 

\begin{equation}
\overline{q}_l(i)=\sqrt{\dfrac{4\pi}{2l+1}\sum_{m=-l}^{l}|\overline{q}_{lm}(i)|^2}
\label{order-parameters-dellago}
\end{equation}

As has been commented previously, the sensibility of the local bond order parameters can be tuned by choosing different values of $l$. The standard approach in order to find the best value of $l$ consists of calculating the local bond order parameters of a bulk liquid phase and a bulk crystal phase at exactly the same thermodynamic conditions. The optimal $l$ value will be the one at which the overlap between the results obtained from the liquid and the crystal phases is minimal, allowing to distinguish accurately which molecules are in the bulk liquid phase and which molecules are in the bulk crystal phase. The percentage of crystal-like molecules labeled as liquid-like molecules and vice versa is called mislabeling.~\cite{Sanz2013a,Espinosa2014b} The value of $\overline{q}_{l}$ from which molecules are labeled as crystal- or liquid-like is chosen in order to have the same mislabeling in both cases. Typically, no higher values of $l=6$ have been employed in the literature for water systems,~\cite{Sanz2013a,Espinosa2014b} although very recently\cite{Grabowska2023a} some of us have used the $\overline{q}_{12}$ parameter to distinguish between water molecules in a liquid CH$_4$ aqueous phase and in a CH$_4$ hydrate phase. In this work, we have analyzed the local bond order parameters of Lechner and Dellago\cite{Lechner2008a} from $l=0$ to $l=20$. Also, it is interesting to notice that it is possible to reduce the mislabeling by using a combination of two $\overline{q}_{l}$. Following this approach, and the idea proposed by Desgranges and Delhommelle,~\cite{Desgranges2008a} the two values of $\overline{q}_{l}$ with the lowest mislabeling values are used in combination to describe a line that separates the crystal-like from the liquid-like molecules. The final mislabeling obtained from the combination of both $\overline{q}_{l}$ is lower than the mislabeling of each $\overline{q}_{l}$ when they are used individually.

Another key value that has to be taken into account is the cutoff distance to identify neighbors. A priory, this value can be chosen arbitrarily, as in the case of $l$, in order to find an optimal value that provides a low mislabeling result. In this work, we have chosen two different cutoff values depending on the system: 0.54 and 0.55 nm. These values correspond to the position of the second minimum of the oxygen–oxygen pair correlation function, $g_{\text{OO}}(r)$, in the hydrate phase. The position of the second minimum shows a poor dependency on the thermodynamic conditions of pressure and temperature, the guest molecule, and the hydrate crystalline structure. As a consequence, despite the different thermodynamic conditions and guests used in this work, the position of the second minimum remains almost constant even for different hydrate structures (sI and sII).

The analysis of local bond order parameters in hydrate systems is slightly different than those involving pure phases, such as ice-water systems. In this latter case, the system has only one component, and each phase in equilibrium is formed from only one substance. According to this, it is necessary to simulate bulk phases of pure substances, i.e., the solid bulk phase (ice) and the pure water liquid bulk phase separately. Following the ice-water example, the trajectories of the molecules of water are extracted from the simulation of both bulk phases. Then, the different $\overline{q}_{l}$ parameters are calculated for each water molecule in the ice and liquid bulk phases. Finally, the mislabeling is calculated for each $\overline{q}_{l}$ to choose the optimal one.

However, if a hydrate phase is involved in the analysis, the system has, at least, two components, water and the guest substance (CO$_{2}$, CH$_{4}$, and other gases). In this case, it is necessary to simulate a bulk phase of hydrate. Usually, we assume full occupancy of all the cages of the hydrate structure by the guest molecules although other occupancies can be considered. The main difference with the case of systems involving pure components is that the bulk liquid phase, which must be simulated as well, is a binary mixture with a given composition. The composition of the guest molecule in the aqueous solution is determined by the coexistence conditions of this phase with the hydrate, which are usually in thermodynamic equilibrium. According to this, it is mandatory to analyze the local bond order parameters from trajectories of simulations in which the composition of the guest molecule in the aqueous phase is exactly that of equilibrium. In other words, following the rule phase, the system has one extra degree of freedom (i.e., the composition of the mixture) that must be taken into account explicitly.

\section{Simulation details and molecular models}

\subsection{Bulk Systems}
In this work, we have studied the capability of the averaged local bond order parameters proposed by Lechner and Dellago\cite{Lechner2008a} to distinguish between liquid-like and hydrate-like water molecules in five different hydrates (CO$_2$, CH$_4$, N$_2$, H$_2$ and THF) at several thermodynamic conditions. In all cases, the trajectories of the different systems under study are generated by performing molecular dynamic simulations using GROMACS (version 4.6.5 in double precision).~\cite{VanDerSpoel2005a} Simulations have been carried out using the  Verlet leapfrog\cite{Cuendet2007a} algorithm with a time step of $2\,\text{fs}$ in the isothermal-isobaric ensemble.~\cite{Allen2017a,Frenkel2002a} In order to fix the temperature, the Nosé-Hoover thermostat\cite{Nose1984a} is applied with a time constant of $2$ ps. The Parrinello-Rahman barostat,~\cite{Parrinello1981a} with a time constant of $2$ ps, is chosen to keep the pressure constant. Since the hydrate structures sI and sII present cubic symmetry, the barostat is applied isotropically in the case of bulk liquid and bulk hydrate systems. The thermodynamic conditions of temperature and pressure at which each system has been studied are specified in Section IV. For the  CO$_2$, CH$_4$, N$_2$, and H$_2$ systems, we use a cut-off of $1.0\,\text{nm}$ for the Coulombic and dispersive interactions. In the case of the THF system, a cut-off of $1.55\,\text{nm}$ for the Coulombic and dispersive interactions is applied. The Fourier term of the Ewald sums is evaluated using the particle mesh Ewald (PME) method~\cite{Essmann1995a} and no long-range corrections are applied for the dispersive interactions. 

CO$_2$ and CH$_4$ hydrates present, under the conditions studied in this work, the sI hydrate structure. In both cases, the bulk hydrate phase is built up by replicating the sI unit cell twice in each space direction ($2\times2\times2$) and assuming single occupancy. The total number of molecules of water and CO$_2$/CH$_4$ in both cases is 368 and 64 respectively. The bulk liquid phases are built up by placing the double molecules of water (736) and the corresponding number of molecules of CO$_2$/CH$_4$ at the thermodynamic conditions under study. In particular, the number of CO$_2$/CH$_4$ molecules in the bulk liquid aqueous phases is calculated from the equilibrium solubility values when the aqueous liquid phase is in contact, via a planar interface, with the corresponding hydrate phase. The solubility values of the CO$_2$ aqueous solutions have been taken from the works of Miguez \emph{et al.}~\cite{Miguez2015a} and Algaba \emph{et al.}~\cite{Algaba2023a} Water molecules are modeled using the TIP4P/Ice model~\cite{Abascal2005b} and CO$_{2}$ molecules are described using the TraPPE model.~\cite{Potoff2001a} Also, The H$_{2}$O--CO$_{2}$ unlike dispersive energy value is given by the modified Berthelot combining rule, $\epsilon_{12}=\xi(\epsilon_{11}\,\epsilon_{22})^{1/2}$, with $\xi=1.13$. The solubility values of the CH$_4$ aqueous solutions are taken from the works of Grabowska \emph{et al.}\cite{Grabowska2022a,Grabowska2023a} In this case, CH$_4$ molecules are described using a spherical Lennard-Jones (LJ) interaction site with the parameters proposed by Guillot and Guissani~\cite{Guillot1993a} and Paschek.~\cite{Paschek2004a} Lorentz-Berthelot rules are used to describe the water-CH$_4$ dispersive cross-interactions. Hydrate and liquid bulk phases of both CO$_2$/CH$_4$ systems have run for $55\,\text{ns}$. The first $5\,\text{ns}$ correspond to the equilibration period and the last $50\,\text{ns}$ to the production period. The trajectories of the water molecules in all cases are collected each $20\,\text{ps}$ over the production period for a total of $2500$ different trajectories of each bulk phase in both systems. The analysis of the $\overline{q}_{l}$ values for the CO$_2$ and CH$_4$ systems is obtained from the analysis of the $2500$ trajectories of each water molecule in each bulk phase. Additionally, we have also performed simulations of a pure water bulk phase. As in the previous case, the pure water system has run in the $NPT$ ensemble for $55\,\text{ns}$ and the trajectories of the water molecules are collected from the last $50\,\text{ns}$ each $20 \,\text{ps}$. As it is mentioned previously, the thermodynamic conditions at which each system is studied are specified in Section IV. 

In this work, we have applied for the first time the averaged local bond order parameters proposed by Lechner and Dellago\cite{Lechner2008a} to three different sII hydrate structures. In particular, we have studied the hydrates of N$_2$, H$_2$, and THF. The hydrate bulk phases are built up by replicating the sII hydrate unit cell twice in each space direction ($2\times2\times$2). We assume single occupancy for the N$_2$ and H$_2$ bulk hydrate phases and the final hydrate simulation box has 1088 and 192 molecules of water and N$_2$/H$_2$ respectively. THF only occupies the H cages (5$^{12}$6$^4$) of the sII hydrate structure (8 per unit cell) while the T cages (5$^{12}$) remain empty (16 per unit cell). The final hydrate simulation box has 1088 and 64 molecules of water and THF respectively. For the three systems, the bulk liquid phases are built up by placing the same number of water molecules as in the hydrate (1088) and the corresponding number of N$_2$, H$_2$, and THF at the thermodynamic conditions used in this work. The solubility of N$_2$ in the aqueous phase is taken from the work of Algaba \emph{et al.}~\cite{Algaba2023b} N$_2$ molecules are modeled through the TraPPE (Transferable Potentials for Phase Equilibria) force field.~\cite{Potoff2001a} Also, unlike dispersive interactions between N$_2$ and water molecules are given by a modification of the Berthelot combining rule, $\epsilon_{12}=\xi(\epsilon_{11}\,\epsilon_{22})^{1/2}$, with $\xi=1.15$. The solubility of H$_2$ in an aqueous phase has been calculated by some of us previously.~\cite{private} H$_2$ molecules are modeled using a modification of the Silvera-Goldman potential.~\cite{Michalis2022a,Silvera1978a,Alavi2005a} As in the work of Michalis \emph{et al.},~\cite{Michalis2022a} the Berthelot combining rule for the unlike dispersive interactions between the molecules of water and H$_2$ is modified by a factor that depends on the temperature (we refer the lector to the original work for further details). Finally, the aqueous bulk liquid phase of THF contains exactly the same number of water and THF molecules as in the hydrate bulk phase since THF hydrate presents an univariant hydrate -- aqueous solution two-phase coexistence curve. Along this univariant two-phase coexistence curve, the composition of both phases is the same.~\cite{Makino2005a,Algaba2024c} THF is modeled as a rigid and planar molecule.~\cite{Garrido2016a,Algaba2018a,Algaba2019a} As in our previous work,~\cite{Algaba2024c} the unlike dispersive interactions between THF and water molecules are given by a modification of the Berthelot combining rule, $\epsilon_{12}=\xi(\epsilon_{11}\,\epsilon_{22})^{1/2}$, with $\xi=1.4$. Hydrate and liquid bulk phases have run for 100 ns. The first 20 ns correspond to the equilibration period and the last 80 ns to the production period. The trajectories of the water molecules were collected each 20 ps over the production period (4000 trajectories) of each bulk system. The values of the averaged local bond order parameters $\overline{q}_{l}$ were obtained from the analysis of the 4000 trajectories of each water molecule in each bulk phase.
 
\subsection{Interfacial Systems}
We have also performed extra simulations to study how the presence of an interface can affect the averaged local bond order parameter values. This is crucial since these parameters are used to distinguish between crystal-like and liquid-like molecules in a system where both types of molecules coexist. Also, in the case of hydrates, we have to take into account not only the hydrate-aqueous interface but also the interface between the aqueous phase and a pure guest phase. In order to study the effect of the interfaces on the local bond order parameters determination, we have focused on the case of the CO$_2$ hydrate at $400\,\text{bar}$ and $287\,\text{K}$. 

First, we have studied the aqueous--CO$_2$ interface in a system with and without a hydrate phase. In order to keep constant the temperature and the pressure, the Nosé-Hoover thermostat\cite{Nose1984a} and the Parrinello-Rahman barostat~\cite{Parrinello1981a} are applied with a time constant of $2\,\text{ps}$. The barostat is only applied in the direction perpendicular to the interface ($z$). The simulation has run $100\,\text{ns}$ in the $NPzT$ ensemble, and the trajectories are analyzed each $20\,\text{ps}$ ($5000$ trajectories in total). Also, we have used the models described previously in the case of the CO$_2$ hydrate and aqueous bulk phases. We have performed simulations of an aqueous phase ($736$ molecules of water) surrounded by two phases of $128$ molecules of CO$_2$, i.e. there are two aqueous-CO$_2$ interfaces present in the system. Also, we have induced the formation of a hydrate phase in the middle of the aqueous phase by placing a mold at the crystallographic position of the molecules of CO$_2$ in the hydrate phase (we refer to the lector to our seminar works\cite{Algaba2022b,Zeron2022a,Romero-Guzman2023a} for further details). When the mold is off, only the two aqueous--CO$_2$ interfaces exist in the simulation box. When the mold is on, it induces the formation of a hydrate phase in the middle of the aqueous phase, creating two additional hydrate-aqueous interfaces in the system.

Finally, we have studied the hydrate-aqueous interface without the presence of an aqueous-guest interface. We have built a simulation box by replicating the CO$_2$ hydrate unit cell four times in each space direction ($4\times4\times4)$. Assuming single occupancy, the final hydrate phase contains 2944 molecules of water and $512$ molecules of CO$_2$. Also, we have added, along the $z$ direction, an aqueous phase with $4000$ molecules of water and 240 molecules of CO$_2$. The number of molecules of CO$_2$ in the aqueous phase is calculated from the value of solubility at $400\,\text{bar}$ and $287\,\text{K}$.~\cite{Algaba2023a} The size of the system is chosen to have an interface large enough to increase the number of water molecules at the hydrate-aqueous interface in order to improve the statistics. Simulations are run in the $NPT$ ensemble and the Parrinello-Rahman barostat~\cite{Parrinello1981a} is applied aniisotropically to avoid any stress from the hydrate solid structure. The system is equilibrated for $2\,\text{ns}$ and the production period is $50\,\text{ns}$. Trajectories are analyzed each $20\,\text{ps}$ (2500 trajectories in total).

\section{Results}

In this section, we present the results obtained in this work using the averaged local bond order parameters to identify water molecules in hydrate phases. We first consider analyzing the order parameters from simulations of one-phase systems. In the second section, we analyze the same order parameters obtained from simulations of systems that exhibit any interface.

\begin{figure}
\includegraphics[width=0.9\columnwidth]{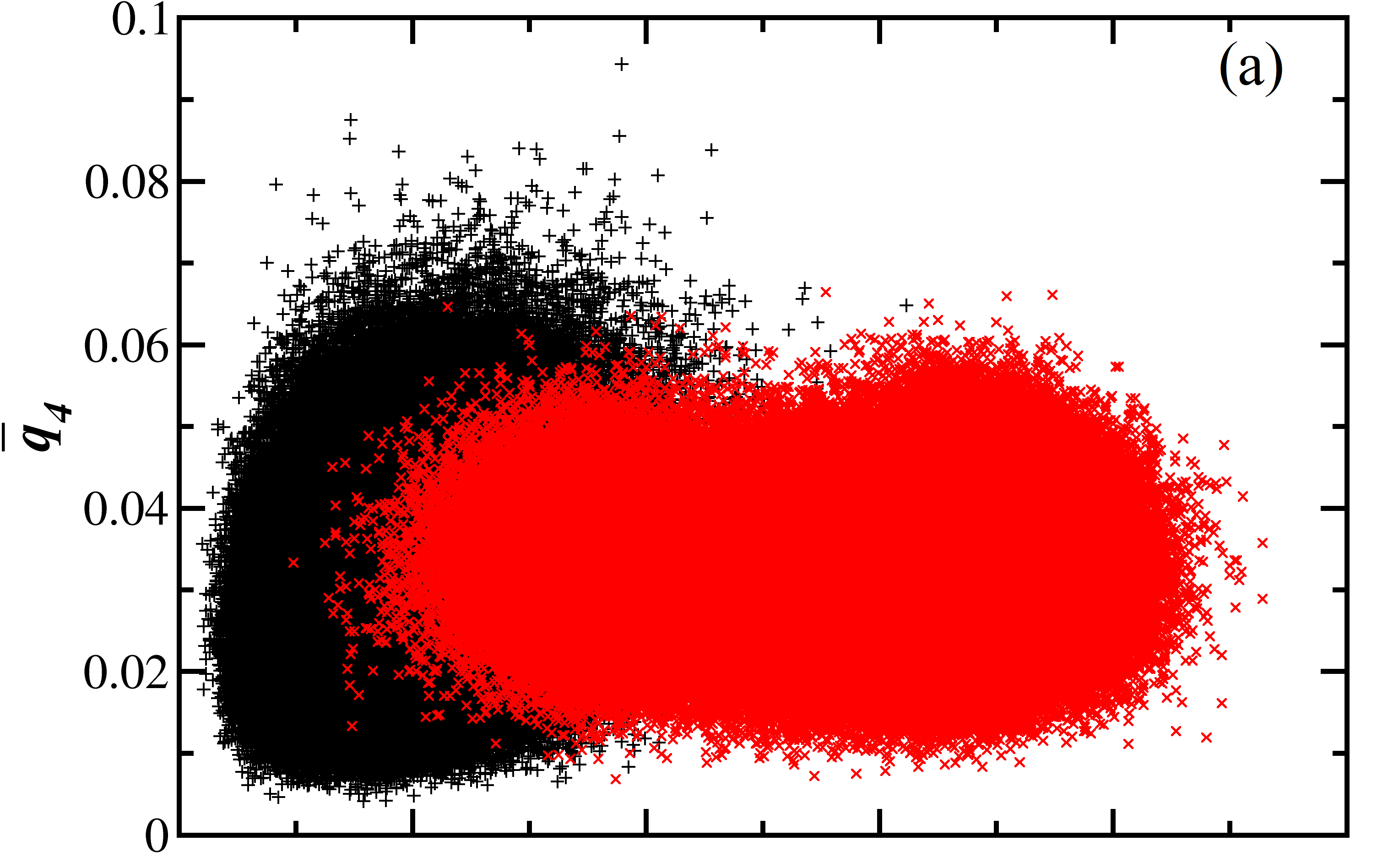}
\includegraphics[width=0.9\columnwidth]{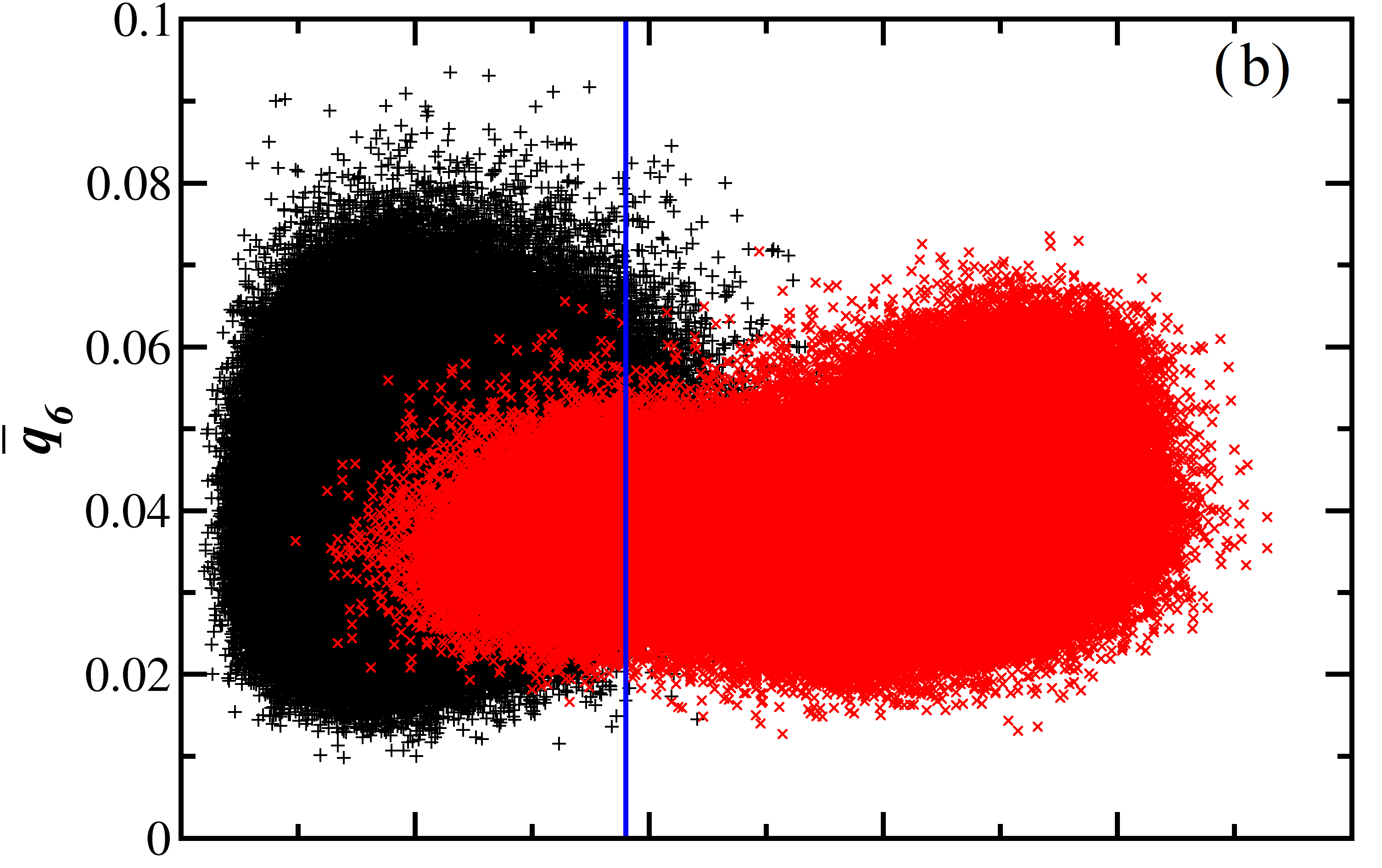}
\includegraphics[width=0.9\columnwidth]{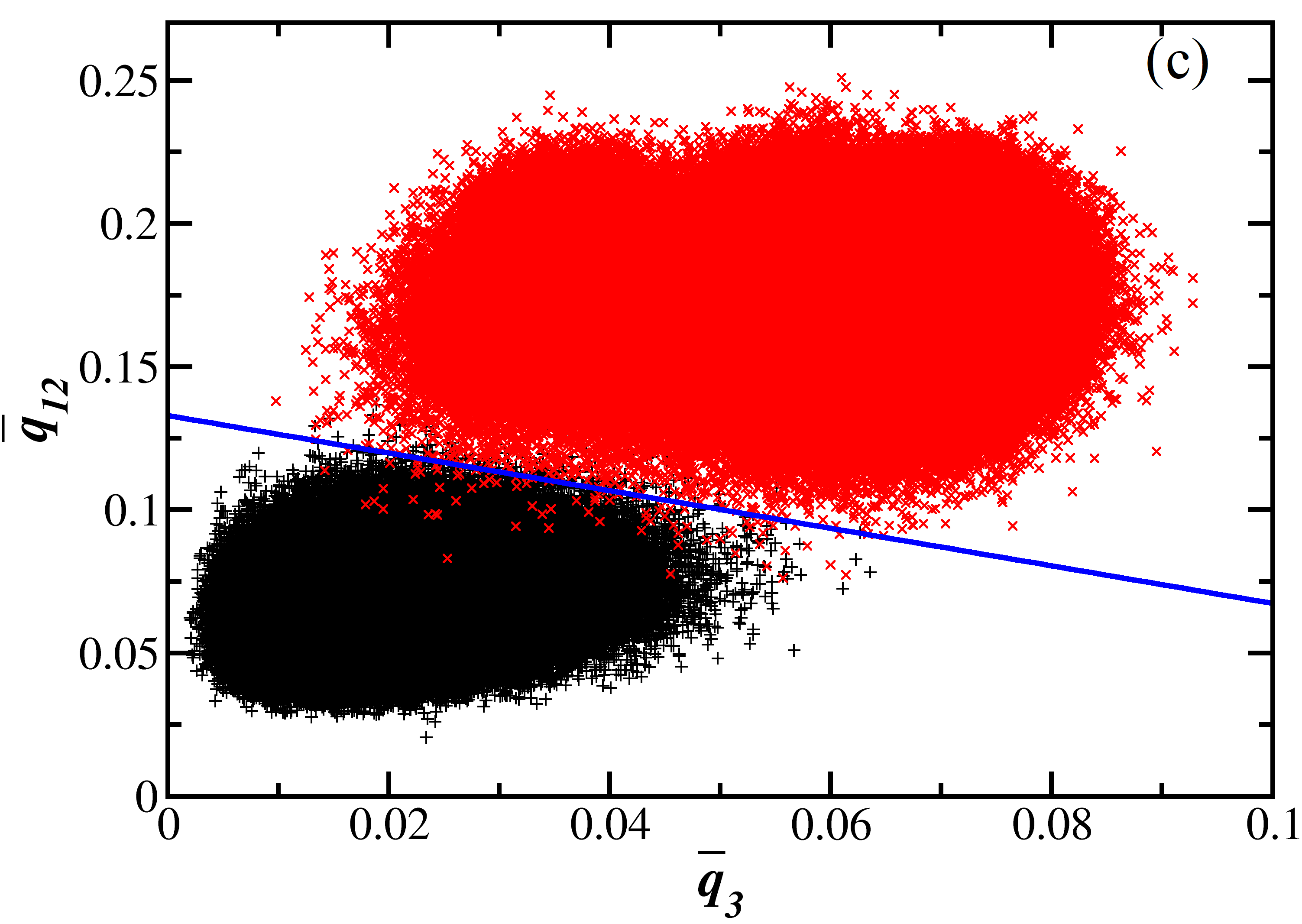}
\caption{Values of (a) $\overline{q}_{4}$, (b) $\overline{q}_{6}$, and (c) $\overline{q}_{12}$, versus $\overline{q}_{3}$, for water molecules at coexistence conditions of the dissociation three-phase lines of the CO$_{2}$ hydrate ($400\,\text{bar}$ and $287\,\text{K}$). Red crosses represent water molecules in the hydrate phase and black pluses to water molecules in the aqueous solution of CO$_{2}$. In all cases, the cutoff distance to identify neighbors is $0.55\,\text{nm}$}
\label{figure1}
\end{figure}

\subsection{Analysis of order parameters from simulation of one-phase systems}

We consider in this section the analysis of local bond order parameters obtained from simulations of two bulk phases: the hydrate crystalline solid phase and the aqueous solution. We concentrate in five different hydrates of CO$_{2}$, CH$_{4}$, N$_{2}$, THF, and H$_{2}$. The two first, CO$_{2}$ and CH$_{4}$ hydrate, are classified as hydrates type sI; the last three exhibit sII crystallographic structure. We analyze these hydrates according to the crystallographic structure exhibited.

\subsubsection{CO$_{2}$ and CH$_{4}$ hydrates (sI structure)}

We first study the case of the CO$_{2}$ hydrate. From the structural point of view, the solid hydrate exhibits the well-known cubic sI crystallographic structure (Pm$\overline{3}$n space group) formed from $64$ water molecules distributed in 2 D (pentagonal dodecahedron or $5^{12}$) cages and 6 T (tetrakaidecahedron or $5^{15}6^{2}$) cages. Assuming single full occupancy, as we have mentioned before, the sI unit cell has 8 additional CO$_{2}$ molecules. From the thermodynamics point of view, this hydrate exhibits a very complex phase diagram. The principal dissociation or three-phase line of the CO$_{2}$ hydrate exhibits two distinct branches: a H--L$_{\text{w}}$--V three-phase line at which the hydrate, the aqueous solution of CO$_{2}$, and the vapor phases coexist, and another three-phase H--L$_{\text{w}}$--L$_{\text{CO}_{2}}$ line where the hydrate, the aqueous solution of CO$_{2}$, and the liquid phase of CO$_{2}$ coexist. Both branches meet at a Q$_{2}$ quadruple point located at $283\,\text{K}$ and $44.99\,\text{bar}$ at which the hydrate, the aqueous solution, the CO$_{2}$ liquid, and the vapor phases coexist. A detailed account of the phases and equilibrium lines of the CO$_{2}$ + water system is explained in terms of its pressure-temperature projection of the phase diagram in the recent work of Algaba \emph{et al.}~\cite{Algaba2022b} We also recommend the reader the books of Sloan and Koh~\cite{Sloan2008a} and Ripmeester and Alavi,~\cite{Ripmeester2022a} for a review of the different structures of hydrates and the associated phase diagrams, and the recent papers of some of us on some useful results for hydrate nucleation, directly related with the calculation of averaged local bond order parameters of CO$_{2}$ hydrates.~\cite{Algaba2022b,Zeron2022a,Algaba2023a,Romero-Guzman2023a}

\begin{figure}[h]
\includegraphics[width=0.9\columnwidth]{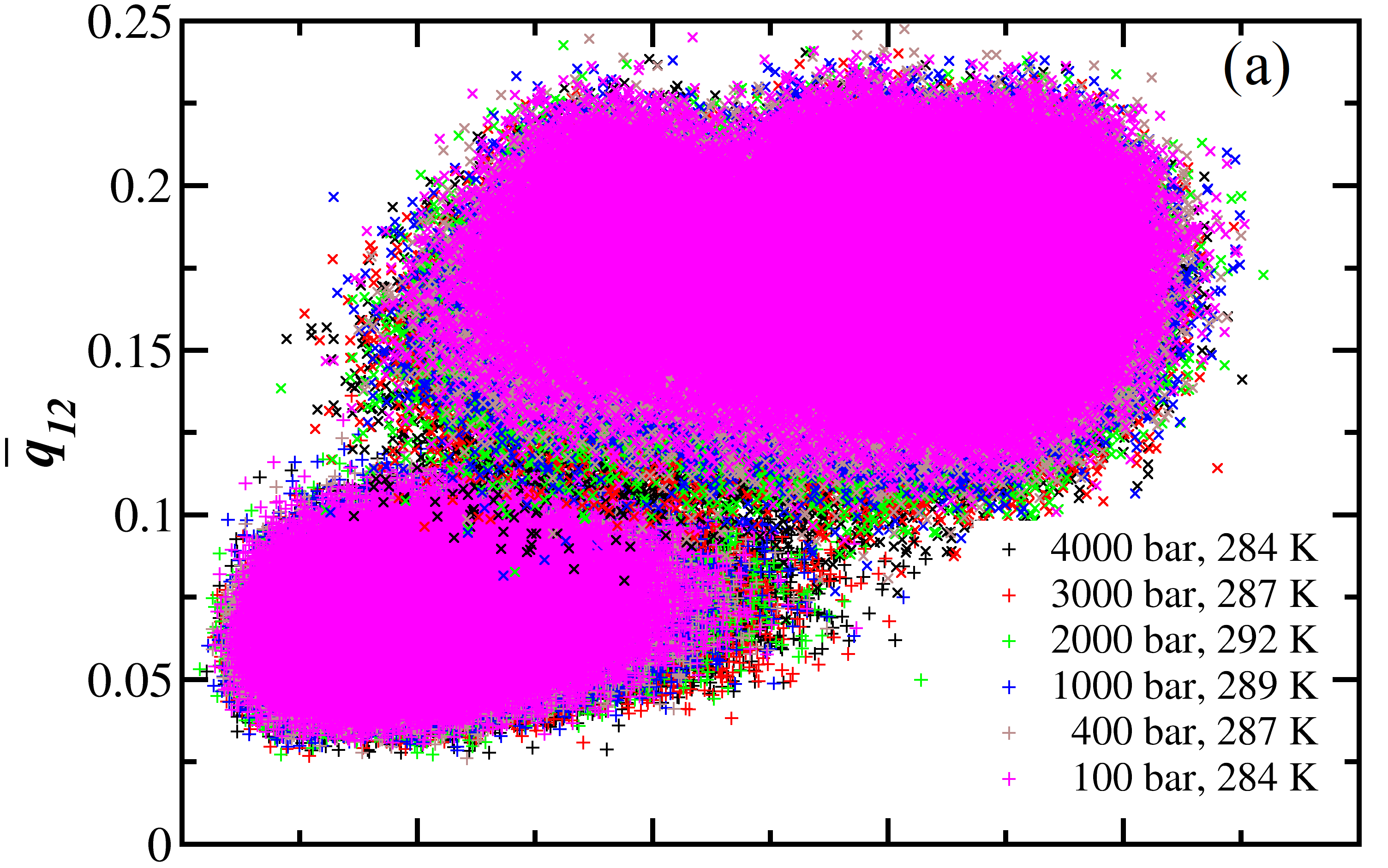}
\includegraphics[width=0.9\columnwidth]{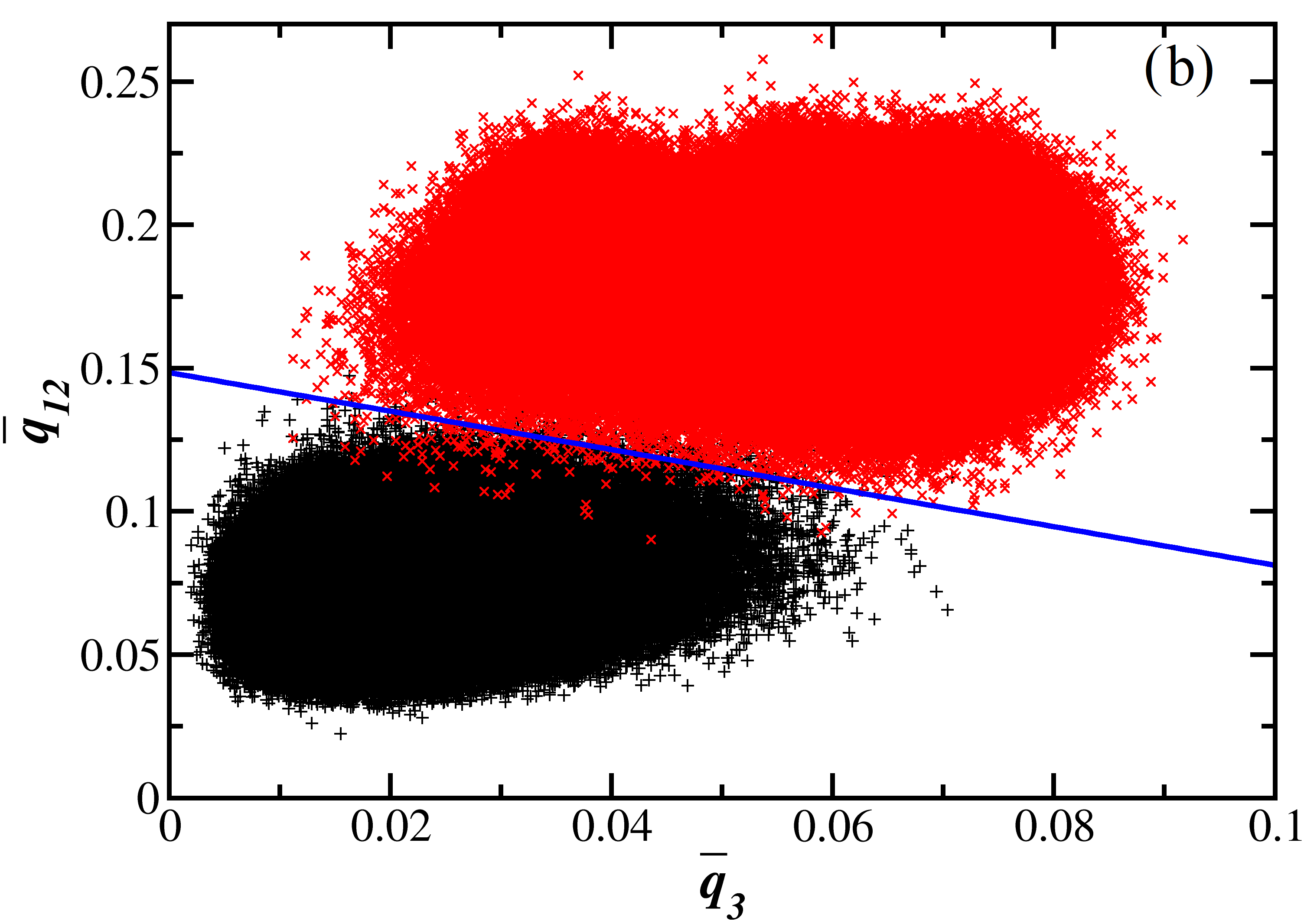}
\caption{Values of $\overline{q}_{12}$ versus $\overline{q}_{3}$, for water molecules (a) at several coexistence conditions along the dissociation three-phase lines of the CO$_{2}$ hydrate and (b) at supercooling conditions $400\,\text{bar}$ and $255\,\text{K}$. In panel (a), different colors represent different conditions of temperature and pressure. The top cloud of points corresponds to water molecules in the hydrate phase and the bottom cloud corresponds to water molecules in the aqueous solution of CO$_{2}$. In panel (b), red crosses represent water molecules in the hydrate phase, and black pluses represent water molecules in the aqueous solution of CO$_{2}$. In all cases, the cutoff distance to identify neighbors is $0.55\,\text{nm}$.}
\label{figure2}
\end{figure}

In this work, we consider the analysis of the behavior and the appropriate election of local-bond order parameter proposed by Lechner and Dellago~\cite{Lechner2008a} that provide the best framework to differentiate if water molecules are in hydrate or aqueous phase along the dissociation three-phase line in which the hydrate, the aqueous solution of CO$_{2}$, and the liquid phase of CO$_{2}$ (H--L$_{\text{w}}$--L$_{\text{CO}_{2}}$) coexist. We have determined the averaged bond order parameters of Lechner and Dellago~\cite{Lechner2008a} at $400\,\text{bar}$ and $287\,\text{K}$. These thermodynamic conditions correspond to a state at the H--L$_{\text{w}}$--L$_{\text{CO}_{2}}$ three-phase line of the CO$_{2}$ hydrate. This state has been previously consider by some of us to determine for the first time the CO$_{2}$ hydrate-water interfacial free energy using the two independent extensions of the Mold Integration technique, namely, the Mold Integration-Host (MI-H)~\cite{Algaba2022b}
and the Mold Integration-Guest (MI-G).~\cite{Zeron2022a} 

We have analyzed several combinations of averaged bond order parameters with different $l$ values. Note that here $l$ is the free integer parameter associated to the first index of the spherical harmonics $Y_{lm}(\mathbf{r}_{ij})$ previously defined. The use of the mislabeling criteria defined previously by Sanz \emph{et al.}~\cite{Sanz2013a} and Espinosa \emph{et al.},~\cite{Espinosa2016c} allows to select the best combination of averaged bond parameters that better describe the separation between the clouds associated to the hydrate and aqueous phases, and consequently, that better identify water molecules as liquid- and hydrate-like.

Fig.~\ref{figure1} shows three representative combinations of averaged bond parameters used previously in the literature~\cite{Lechner2008a,Sanz2013a} and by some of us in previous works.~\cite{Algaba2022b,Zeron2022a,Romero-Guzman2023a} The first one, shown in Fig.~\ref{figure1}a, corresponds to the combination $\overline{q}_{4}-\overline{q}_{3}$. This combination has been used to distinguish different crystalline structures of systems formed from molecules that interact via the hard-sphere and Lennard-Jones intermolecular potentials. Particularly, Lechner and Dellago show that the combination $\overline{q}_{4}-\overline{q}_{3}$ allows to differentiate between body-centered cubic (bcc), face-centered cubic (fcc), and (hexagonal close-packed) hcp crystalline structures and liquid phase of the LJ system at supercooling conditions.~\cite{Lechner2008a} Sanz \emph{et al.}~\cite{Sanz2013a} and Espinosa \emph{et al.}~\cite{Espinosa2014b} have also used this combination successfully to distinguish ice- and liquid-like water molecules in the context of the Seeding technique to estimate homogeneous ice nucleation rates for several models of water. Unfortunately, this combination fails in the case of the CO$_{2}$ hydrate. As can be seen in panel (a) of Figure \ref{figure1}, the clouds associated with water molecules in the hydrate phase (red) and to water molecules in the aqueous solution of CO$_{2}$ (red) overlap in a wide range of $\overline{q}_{3}$ and $\overline{q}_{4}$ values. Particularly, the mislabeling associated to this selection is $2.04\,\%$. Although this value can be used to identify if water molecules are in the liquid or hydrate phase, it is not the most appropriate election as we will show in the next paragraphs.

It is possible to obtain a larger separation between the clouds associated with water molecules in the liquid and hydrate phases increasing the index $l$ associated with the spherical harmonics. Fig.~1b shows the combination $\overline{q}_{6}-\overline{q}_{3}$ values for the same system and at the same thermodynamic conditions. Although it seems that overlapping between the clouds associated with liquid (black pluses) and hydrate (red crosses) is similar, in this case, it is possible to define a threshold value for $\overline{q}_{3}$, $\overline{q}_{3,t}$, that effective separates the two clouds with a more acceptable mislabeling value. Particularly, and according to the results presented in panel (b), water molecules with $\overline{q}_{3}$ values below $\overline{q}_{3,t}\approx 0.038$ are considered liquid-like and those with values above this value hydrate-like. In this case, the mislabeling is equal to $2.03\%$. This combination of averaged bond parameters has been effectively utilized by some of us to determine the number of water molecules in the hydrate phase using the Mold Integration method for hydrates. This approach allowed us to calculate, for the first time, the CO$_{2}$ hydrate -- water interfacial free energy from computer simulations~\cite{Algaba2022b,Zeron2022a,Romero-Guzman2023a}

\begin{figure}[h]
\includegraphics[width=0.85\columnwidth]{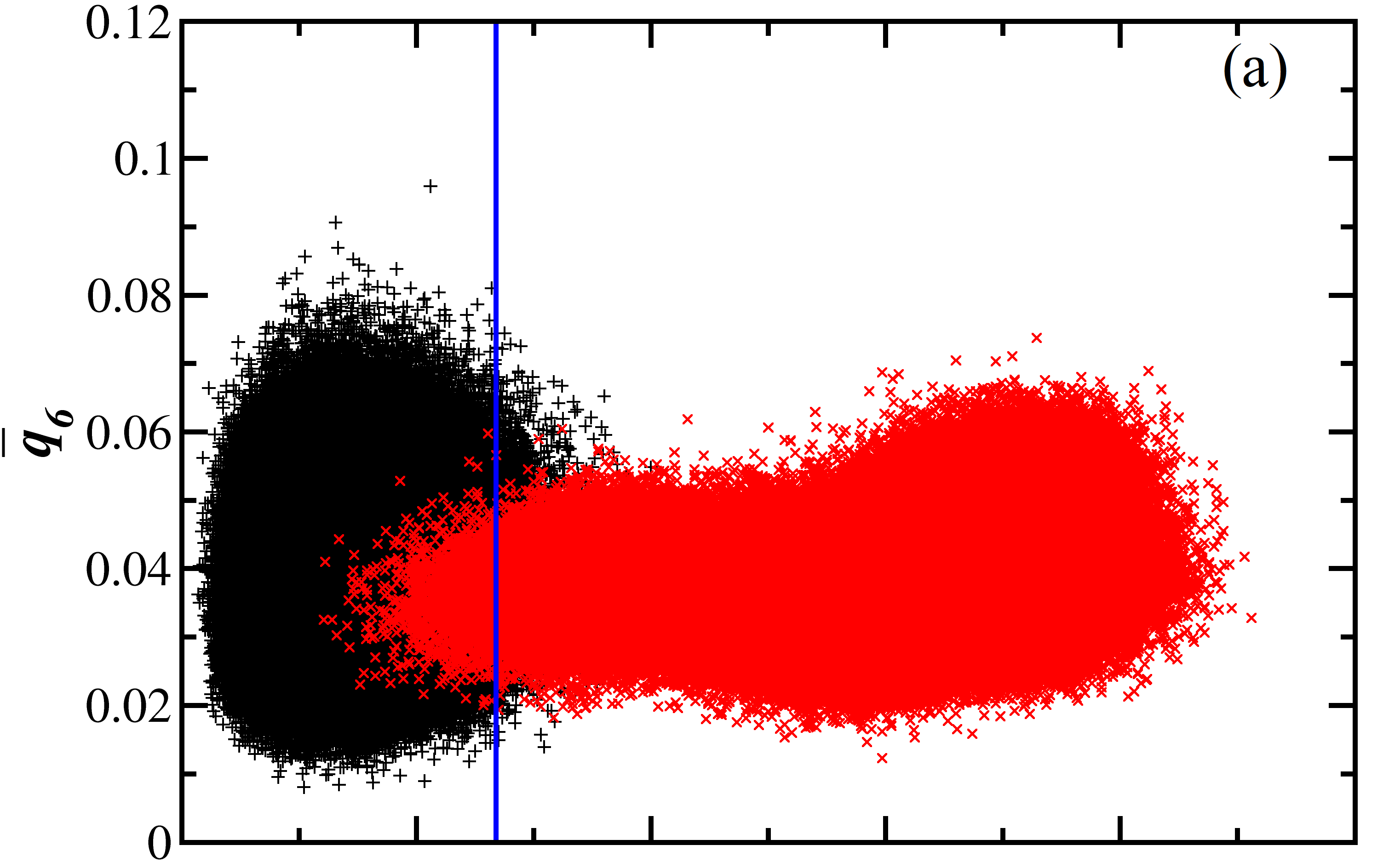} 
\includegraphics[width=0.85\columnwidth]{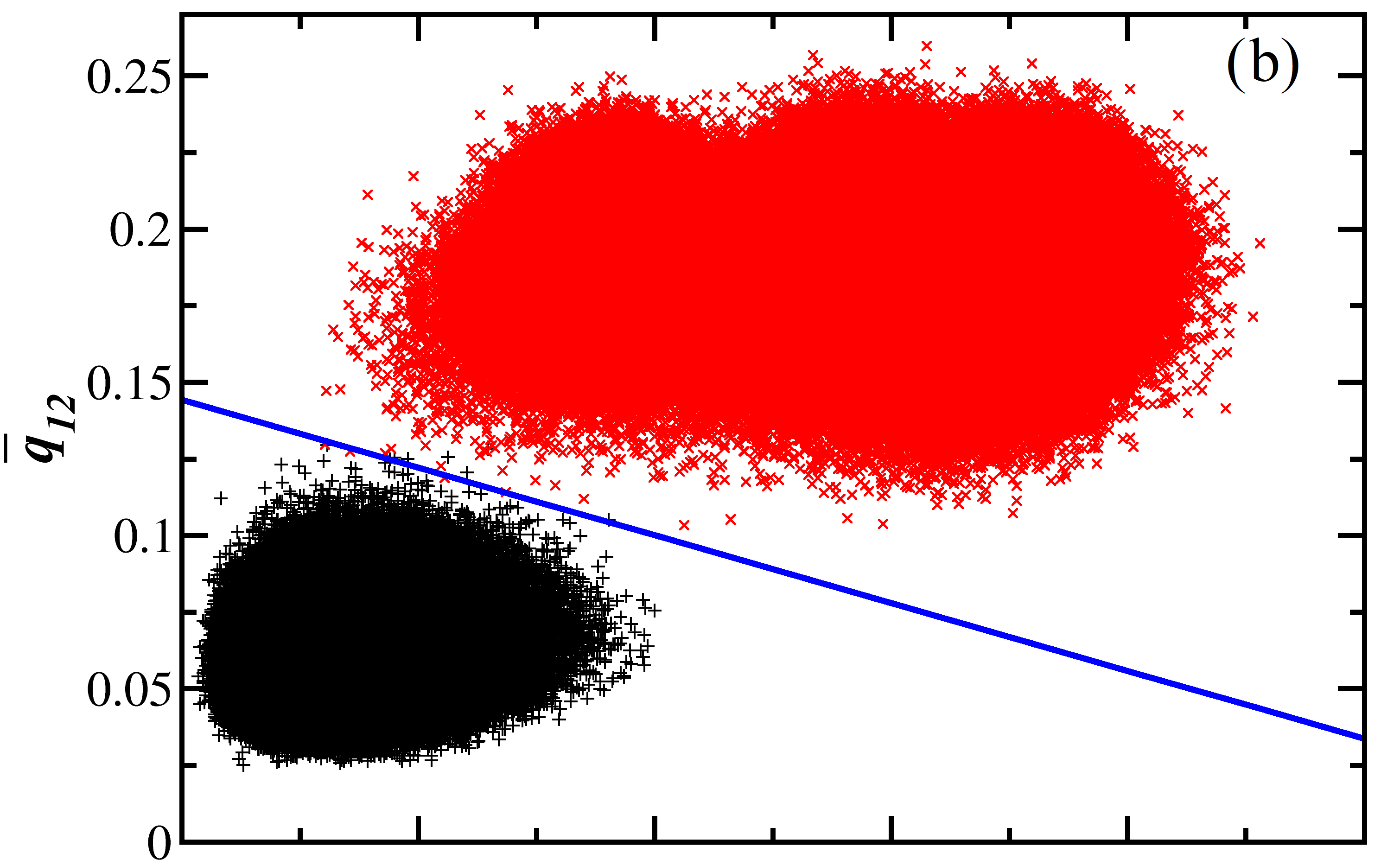}
\includegraphics[width=0.85\columnwidth]{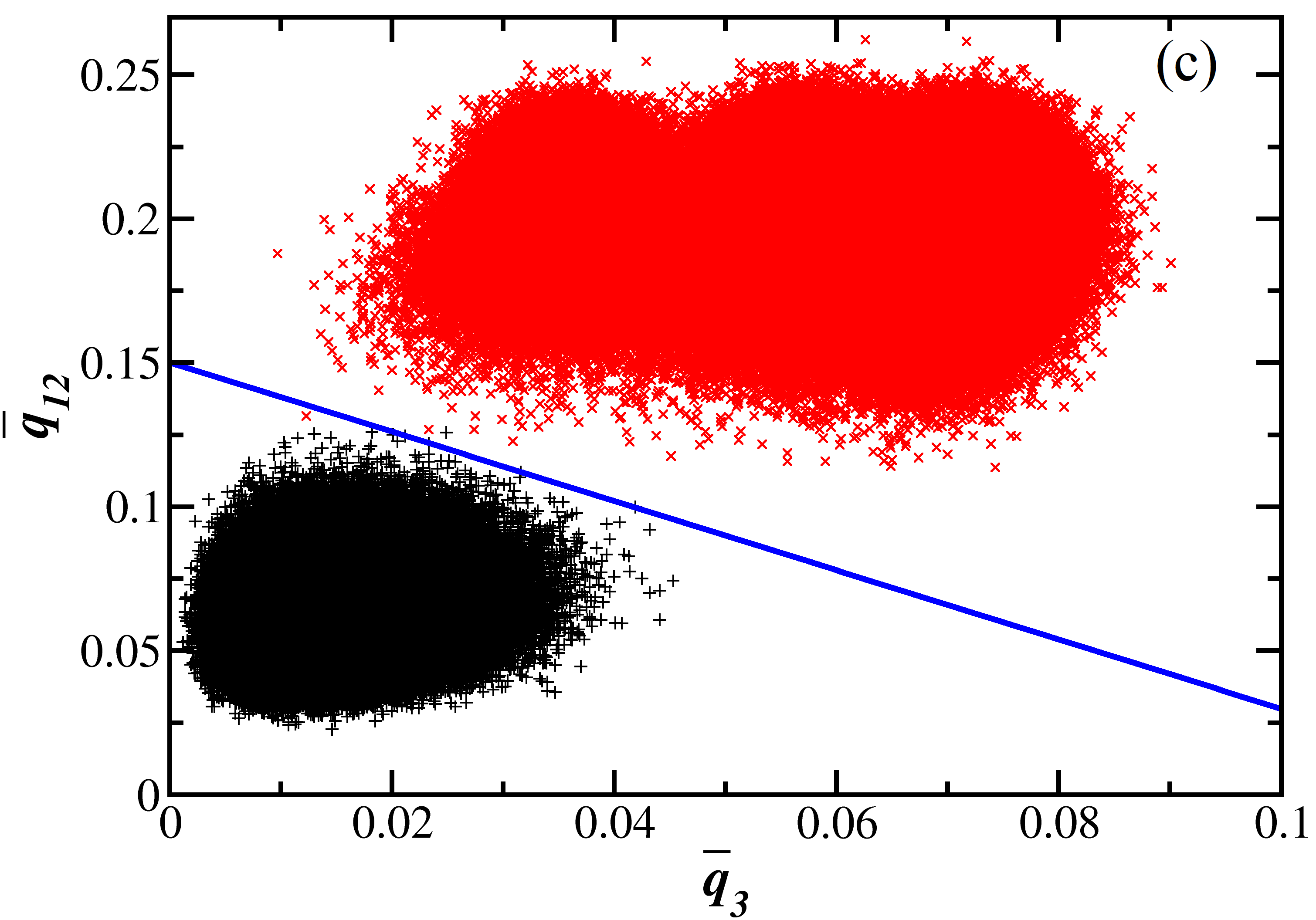}
\caption{Values of (a) $\overline{q}_{6}$ and (b) $\overline{q}_{12}$, versus $\overline{q}_{3}$, for water molecules at coexistence conditions of the dissociation three-phase lines of the CH$_{4}$ hydrate ($400\,\text{bar}$ and $295\,\text{K}$). In panel (c),
$\overline{q}_{12}$ is represented as a function of  $\overline{q}_{3}$ at supercooling conditions $400\,\text{bar}$ and $260\,\text{K}$. In all cases, the top cloud of points corresponds to water molecules in the hydrate phase (red crosses) and the bottom cloud to water molecules in the aqueous solution of CH$_{4}$ (black pluses). In all cases, the cutoff distance to identify neighbors is $0.55\,\text{nm}$.} 
\label{figure3}
\end{figure}

In this work, we have gone further and extended the search of combinations of averaged bond parameters that optimize the separation of clouds associated to liquid-like and hydrate-like water molecules. To this end, we have analyzed the values of the averaged bond order parameter from $l=1$ to $20$. The most important and relevant result of this analysis is that the best combination of all the $170$ possible $\overline{q}_{i}$--$\overline{q}_{j}$ different pairs (with $i,j=1, \ldots ,20$ and $i\ne j$) considered in this work is that in which one of the averaged bond parameters is $\overline{q}_{12}$. Particularly, the mislabeling found runs from $0.0117$ and $0.0646\%$. The largest mislabeling obtained using a combination of two averaged bond parameters, being one of them $\overline{q}_{12}$ (at $400\,\text{bar}$ and $287\,\text{K}$), is $\overline{q}_{12}-\overline{q}_{7}$. Fig.~\ref{figure1}c shows the combination of the $\overline{q}_{12}$ and $\overline{q}_{3}$
averaged bond parameters that provide, according to our study, the largest separation between the clouds associated with liquid-like and hydrate-like water molecules. As can be seen, it is possible to define a threshold value for $\overline{q}_{12}$, $\overline{q}_{12,t}\approx 0.12$, that separates very well the two clouds with a more acceptable mislabeling value, better than in the cases presented in panels (a) and (b). However, we have opted to choose as threshold a linear combination of the values of the averaged order parameters $\overline{q}_{12}$ and $\overline{q}_{3}$. The reason is clear: with this election, the accuracy of the assignment is clearly improved since the mislabeling is $0.0117\%$. This election has been used by some of us to calculate nucleation rates
of CO$_{2}$ hydrates from seeding runs in solutions of CO$_{2}$ in water.~\cite{Zeron2024a}

As we have already mentioned, local order parameters are routinely used to identify if molecules are in a liquid or a solid phase. For instance, this happens by applying the seeding technique in pure systems,~\cite{Sanz2013a} where the order parameters are used to estimate the size of the emerging solid cluster of molecules in the metastable phase. Accurate identification and counting of molecules in the solid phase is the key point of the methodology since the size of the cluster critically depends on this identification, and consequently, the estimated homogeneous nucleation rate. In nucleation, the emerging solid cluster and the metastable liquid phase are in equilibrium. If thermodynamic conditions are changed, i.e., temperature or pressure, equilibrium conditions also vary and this affects the topological representation of the order parameters: the molecular structure of the solid changes but the structure of the liquid changes as well. This provokes a variation of the threshold already established and could introduce spurious bias for the correct identification of molecules in the solid phase. The final result is an erroneous estimation of the size of the emerging solid cluster in the metastable phase.

The scenario is similar in the case of nucleation of hydrates but with an important difference: hydrates are by nature mixture systems. This means that composition, the new thermodynamic degree of freedom that enters according to the phase rule of Gibbs, play also a key role. It is important to note that the solubility of CO$_{2}$ in water changes with temperature and pressure and this affects the molecular structure and composition of the aqueous solution of CO$_{2}$ and consequently the topology of the $\overline{q}_{12}-\overline{q}_{3}$ representation. In this context, an interesting question arises: are the averaged order parameters used at a given thermodynamic conditions valid at other conditions, at which composition varies significantly? To clarify this important point, we have extended the conditions at which the order parameters have been presented. To this end, we consider two different situations. The first one is the analysis of the $\overline{q}_{3}$ and $\overline{q}_{12}$ averaged order parameters at different temperatures and pressures along the dissociation H--L$_{\text{w}}$--L$_{\text{CO}_{2}}$ three-phase line of the CO$_{2}$. The second one is the evaluation of the same order parameters at supercooling conditions at which homogeneous nucleation rates could be determined.

Figure~\ref{figure2} shows the $\overline{q}_{12}-\overline{q}_{3}$ representation at thermodynamic states along the dissociation line of the CO$_{2}$ hydrate, from $100$ to $4000\,\text{bar}$. As can be seen in panel (a), the representation shows two compact clouds associated to water molecules in the hydrate phase (top cloud) and the aqueous solution of CO$_{2}$ (bottom) well separated. Particularly, the mislabeling at the pressures presented in panel (b), from $100$ to $4000\,\text{bar}$, are $0.0102$, $0.0117$, $0.0204$, $0.0252$, $0.0361$, and $0.0462\%$, respectively. There is a slight increase in the mislabeling as the pressure increases. Since the dissociation H--L$_{\text{w}}$--L$_{\text{CO}_{2}}$ three-phase line of the CO$_{2}$ exhibits a positive slope at this range of pressure, the temperature also increased, provoking more thermal noise in the system making a slightly more difficult to distinguish between liquid-like and hydrate-like water molecules. However, at this point is important to remark, that the threshold used in Fig.~\ref{figure2}b corresponds to that chosen at 400 bar. This is relevant because the mislabeling obtained at 100 bar is lower than that obtained at 400 bar. Also, as can be observed in Fig.~\ref{figure2}b, the core of each bulk phase cloud remains constant. It means that, at least in this case, the same $\overline{q}_{12}-\overline{q}_{3}$ representation can be used to analyze the CO$_2$ hydrate - aqueous system along the three-phase coexistence line, and the increase of the mislabeling with the increase of pressure can be related exclusively to the increase of the thermal noise.

We have also extended the study and tried to use the $\overline{q}_{12}-\overline{q}_{3}$ representation in a different thermodynamic state. This state corresponds to the usual situation at which nucleation rates of hydrates and other solid phases are evaluated: a state at which the temperature is below the equilibrium thermodynamic between two or more phases. As we have already stated, at $400\,\text{bar}$, the molecular models describing the CO$_{2}$ hydrate exhibit a three-phase equilibrium between the hydrate, the aqueous solution of CO$_{2}$, and a CO$_{2}$-rich liquid phase at $290\,\text{K}$.~\cite{Algaba2023a} Note that the value of dissociation temperature of the CO$_{2}$ hydrate obtained in the first estimation by M\'{\i}guez \emph{et al.}~\cite{Miguez2015a} was $287(2)\,\text{K}$. Recently, some of us~\cite{Algaba2023a} have obtained a slightly higher value using a larger system and cut-off distance, $290(2)\,\text{K}$. Both results are within the error bars and differences between order parameters obtained using these two temperature values are negligible. In this particular case we choose the state at $400\,\text{bar}$ and $255\,\text{K}$. This corresponds to a supercooling of $35\,\text{K}$. Fig.~\ref{figure2}b shows the $\overline{q}_{12}-\overline{q}_{3}$ representation at this state. As can be seen, the two clouds associated with liquid-like and hydrate-like water molecules are separated using the linear combination of the values of the averaged order parameters $\overline{q}_{12}$ and $\overline{q}_{3}$ previously used. In this case, the mislabeling also corresponds to a very low value of $0.0183\%$. 

We have already checked that the appropriate combination of averaged order parameters allows us to identify easily liquid-like and hydrate-like water molecules for the CO$_{2}$ hydrates in different situations. We now turn on to the study of other hydrate, that exhibits the same sI crystallographic structure, the CH$_{4}$ methane. Although the solid structure is the same, i.e., 64 water molecules that form 2 D and 6 T cages, with the CH$_{4}$ molecules occupying both the large (T) and small (D) cages, the phase diagram of this hydrate, and particularly its $PT$ projection, is topologically different from a thermodynamic point of view. Since CH$_{4}$ has a critical temperature at very low temperatures ($T_{c}\approx190\,\text{K}$), the CH$_{4}$ hydrate only exhibits a unique quadruple point and the second characteristic quadruple point (Q$_{2}$) present in other hydrates, including the CO$_{2}$ hydrate, is absent. The reason is that no liquid phase for CH$_{4}$ exists at temperatures at which the hydrate forms. This quadruple point is usually absent when the guest is a light gas like methane. As we have already mentioned, the solid phases of the CO$_{2}$ and CH$_{4}$ hydrates are similar from the structural point of view. However, the structural distribution of water molecules in the aqueous phase is different in both systems due to the specific CO$_{2}$--water and CH$_{4}$--water molecular interactions. Particularly, the solubility of CH$_{4}$ in water is much lower than that of CO$_{2}$ (ten times lower, approximately). This effect can change the distribution of the bond order parameters in the $\overline{q}_{12}-\overline{q}_{3}$ representation affecting the distinguishability between liquid-like and solid-like in hydrate systems. 

Figure~\ref{figure3} shows the $\overline{q}_{6}-\overline{q}_{3}$  and $\overline{q}_{12}-\overline{q}_{3}$ representations, at several thermodynamic conditions, of water molecules in aqueous solutions of CH$_{4}$ and the CH$_{4}$ hydrate solid phase. In panels (a) and (b) we present the bond order parameters, at $400\,\text{bar}$ and $295\,\text{K}$. These thermodynamic conditions correspond to a state at the dissociation line or H--L$_{\text{w}}$--V three-phase line of the CH$_{4}$ hydrate previously studied by some of us.~\cite{Grabowska2022a,Grabowska2023a} As can be seen, there is an important overlap of the clouds associated with the liquid-like (black pluses) and the hydrate-like (red crosses) water molecules in panel (a) ($\overline{q}_{6}-\overline{q}_{3}$ representation). In this case, the mislabeling is $0.3\%$. Although this is an acceptable value to distinguish between water molecules in both phases, it is not the best option. According to panel (b), the $\overline{q}_{12}$ and $\overline{q}_{3}$ combination exhibits better segregation of water molecules in this plane. The most conservative option, not shown in the figure, is to choose only a threshold value for the bond order  $\overline{q}_{3}$. The value that minimizes the mislabeling for both, liquid-like and hydrate-like water molecules is $\overline{q}_{3,t}\approx 0.297\%$. As in the case of the CO$_{2}$ hydrate, it is possible to obtain a better separation between the two clouds using a linear combination of the averaged order parameters $\overline{q}_{12}$ and $\overline{q}_{3}$. This option, shown in panel (b), maximizes the separation between the two clouds, minimizes the mislabeling, $0.0003\%$, and allows to differentiate with confidence liquid-like and hydrate-like water molecules for the CH$_{4}$ hydrate.

We have also used the same averaged bond order parameters to differentiate liquid-like and hydrate-like water molecules at supercooling conditions. In this case, we have chosen a supercooled stated at $400\,\text{bar}$ and $260\,\text{bar}$. Some of us have already published the use of these order parameters successfully to investigate the homogeneous nucleation rate of CH$_{4}$ hydrate formation under experimental conditions using the Seeding simulation technique.~\cite{Grabowska2023a} As can be seen, the
same combination can also provide a good description of the separation of liquid-like and hydrate-like water molecules at supercooling conditions. Particularly, the value of the corresponding mislabeling is now $0.00011\%$. Note that the mislabeling decreases as the temperature is decreased, from  $295$, shown in panel (b), to $260\,\text{K}$, shown in panel (c). This is an expected behavior since the size of the clouds decreases as the thermal energy available for water and guest molecules also decreases with temperature. 

Finally, it is also interesting to compare the $\overline{q}_{12}-\overline{q}_{3}$ representations of the CO$_{2}$ and CH$_{4}$ hydrates at coexistence conditions, shown Figs.~\ref{figure1}c and \ref{figure3}b, and at supercooling conditions, shown Figs.~\ref{figure2}b and \ref{figure3}c, respectively. Note that the pressure is the same in all cases, $400\,\text{bar}$. Coexistence temperatures are slightly different, $287$ and $290\,\text{K}$, but the plots at the metastable states present the same supercooled temperature, $\Delta T=35\,\text{K}$. In the first comparison, considering hydrates at coexistence conditions, the mislabeling decreases from $0.0117\%$ (CO$_{2}$ hydrate) to $0.0003\%$ (CH$_{4}$ hydrate). In the second one, the mislabeling also decreases from $0.0183\%$ (CO$_{2}$ hydrate) to $0.00011\%$ (CH$_{4}$ hydrate). Why does the mislabeling decrease when changing from the CO$_{2}$ to the CH$_{4}$ hydrate? The reason is the high solubility of CO$_{2}$ in water compared with that of CH$_{4}$ in water. The solubility of methane in water is so low that the aqueous solution can considered very dilute.~\cite{Grabowska2022a} Under these conditions, the local environment of water molecules is similar to that of pure liquid water, and at the same time, very different from that of water molecules in the hydrate phase.  The solubility of CO$_{2}$ in water is not negligible,~\cite{Algaba2023a} and the presence of CO$_{2}$ molecules in the aqueous phase modify the local environment of water molecules, increasing the size of the cloud associated to the liquid-like molecules.


\subsubsection{N$_{2}$, THF, and H$_{2}$ hydrates (sII structure)}

We now extend the applicability of the local bond order parameters of Lechner and Dellago to deal with hydrates that exhibit sII crystallographic structure. In this context, an important question arises. Are the selection of the order parameters of Lechner and Dellago~\cite{Lechner2008a} and particularly, is the $\overline{q}_{12}-\overline{q}_{3}$ representation able to properly distinguish between liquid-like and hydrate-like water molecules of hydrates of type sII? To answer this important question, we have considered a reduced but selected number of hydrates that exhibit this structure: nitrogen (N$_{2}$), hydrogen (H$_{2}$), and tetrahydrofuran (THF) hydrates.

\begin{figure}
\includegraphics[width=0.9\columnwidth]{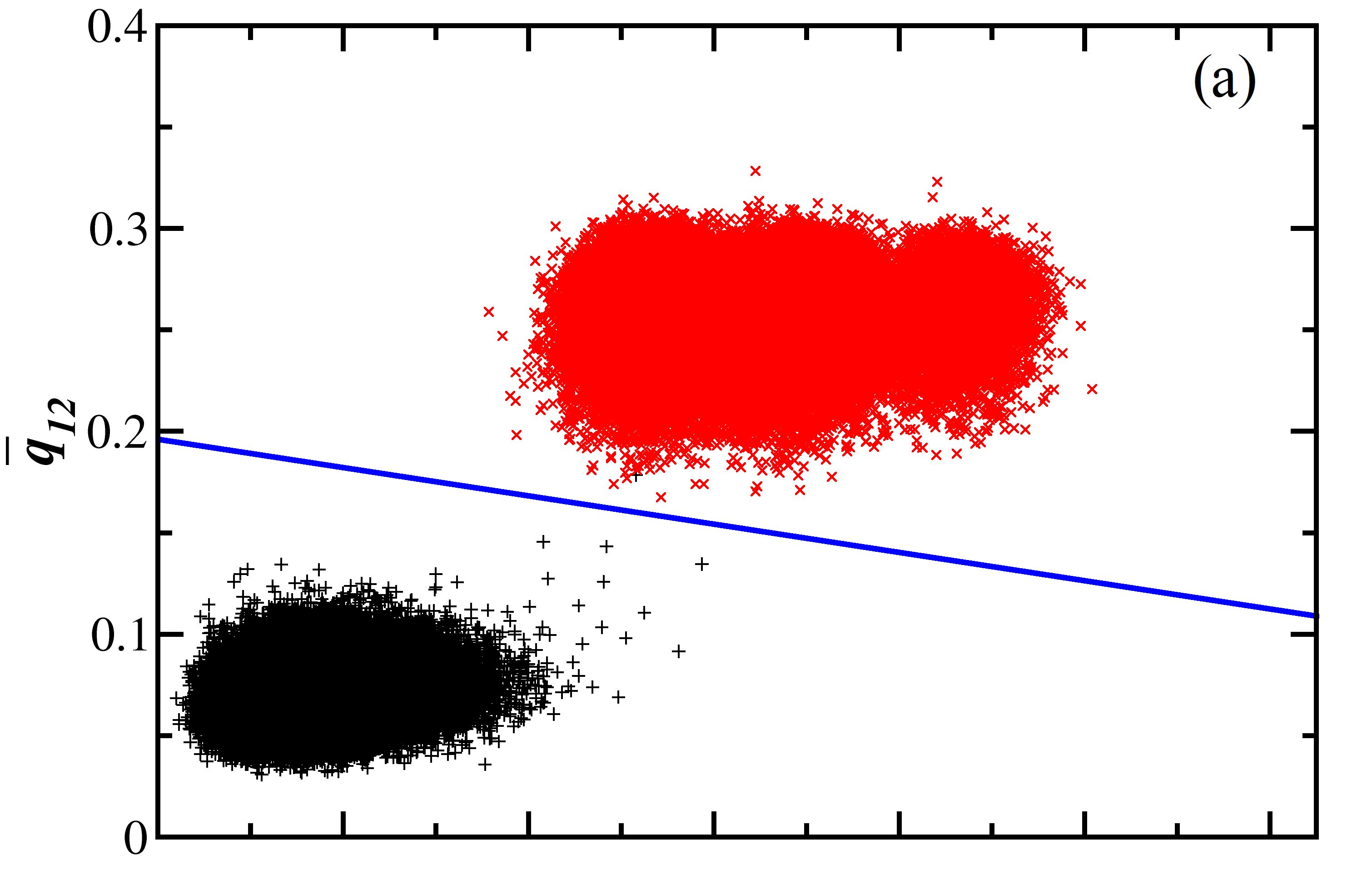}
\includegraphics[width=0.9\columnwidth]{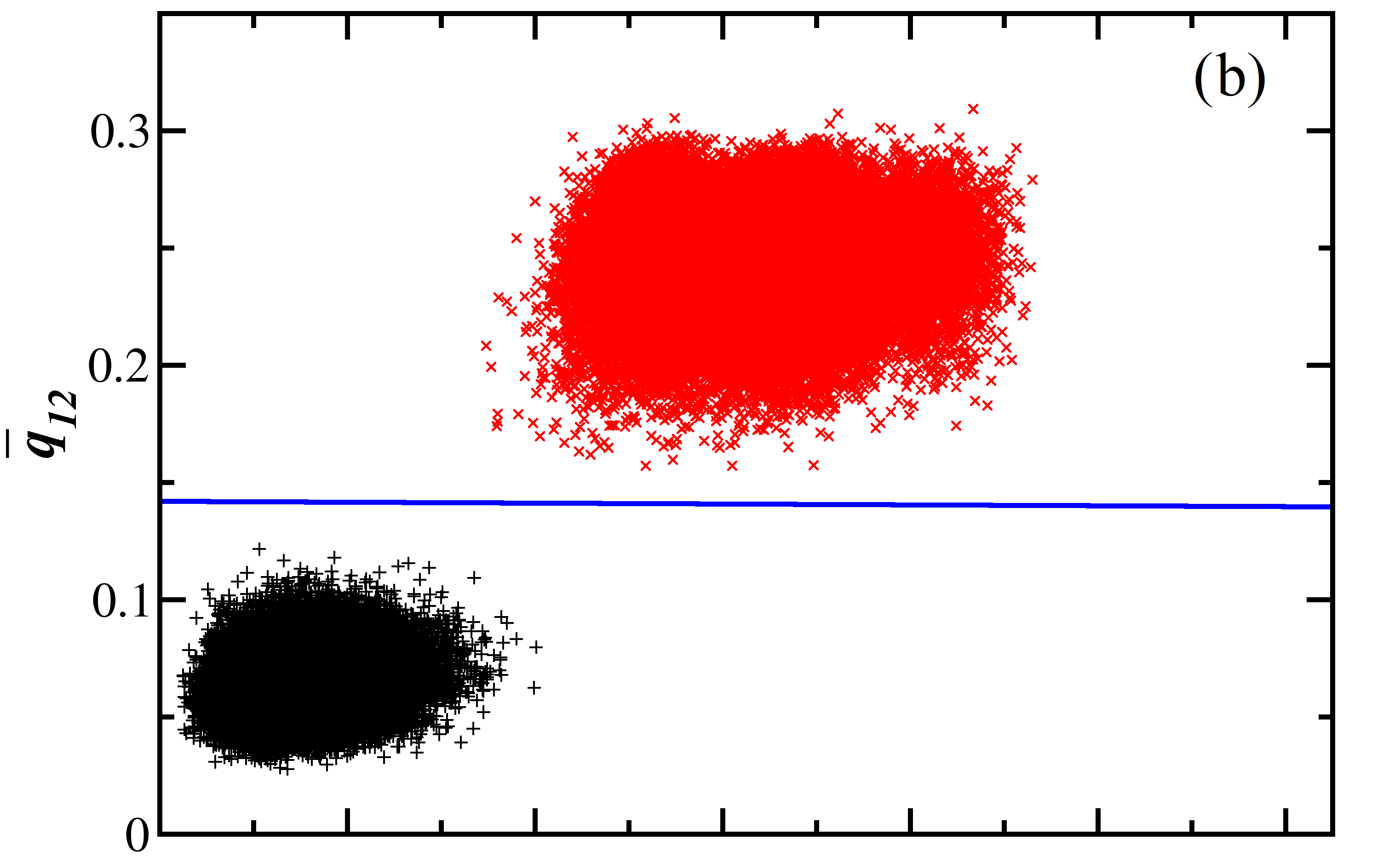}
\includegraphics[width=0.9\columnwidth]{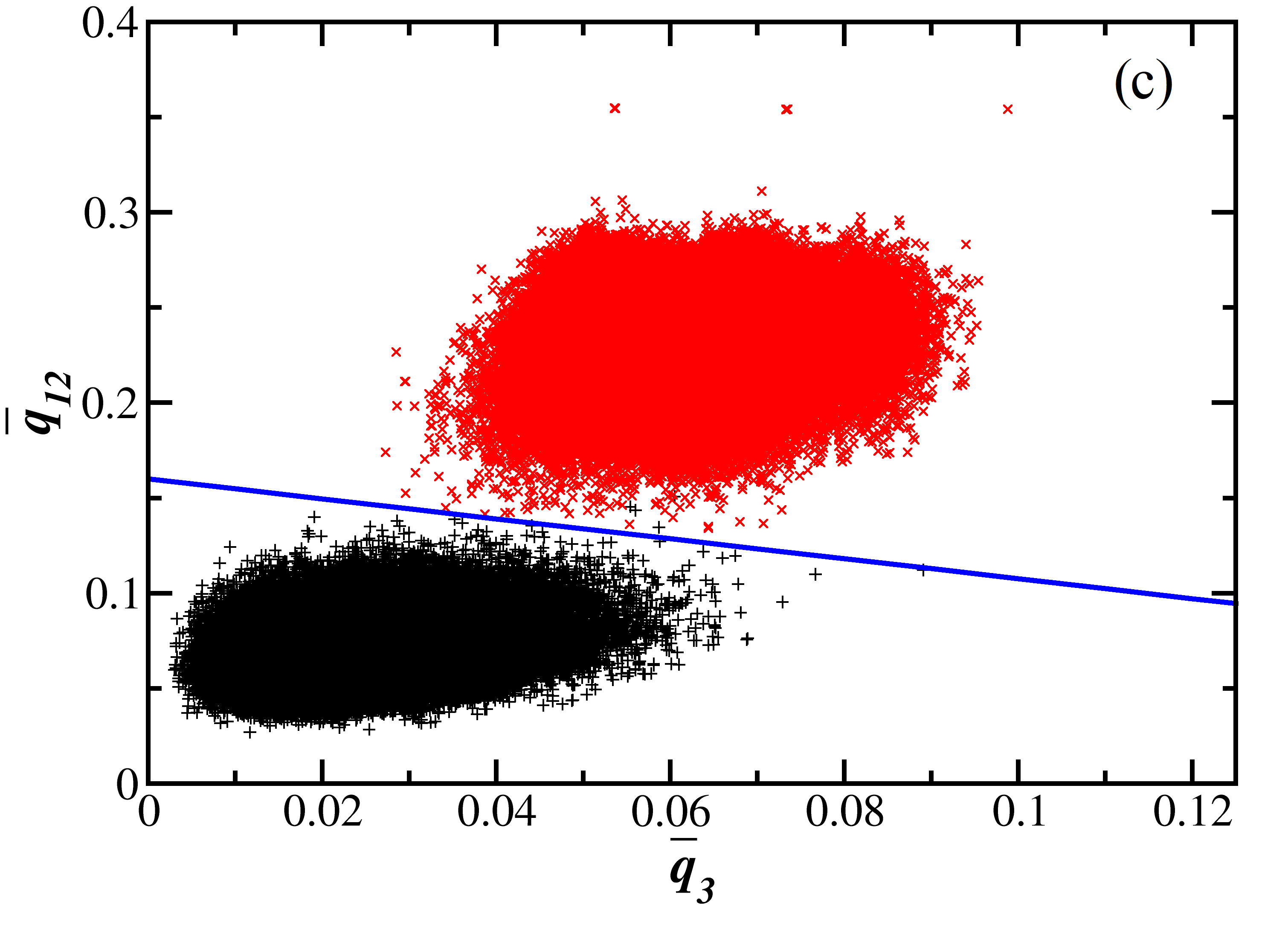}
\caption{Values of $\overline{q}_{12}$ versus $\overline{q}_{3}$, for water molecules at coexistence conditions of the dissociation lines of the (a) N$_{2}$ hydrate at $500\,\text{bar}$ and $250\,\text{K}$ (supercooling of $35\,\text{K}$), (b) H$_{2}$ hydrate at $1850\,\text{bar}$ and $240\,\text{K}$ (supercooling of $35\,\text{K}$), and (c) THF hydrate at $500\,\text{bar}$ and $272.5\,\text{K}$ (two-phase coexistence). In all cases, the top clouds of points correspond to water molecules in the hydrate phase (red crosses) and the bottom clouds to water molecules in the aqueous solution of N$_{2}$, H$_{2}$, and THF (black pluses).}
\label{figure4}
\end{figure}

From the structural point of view, the three hydrates exhibit the well-known cubic sII crystallographic structure (Fd3m space group) formed from $136$ water molecules distributed in 16 D (pentagonal dodecahedron or $5^{12}$) cages and 8 H (hexakaidecahedron or $5^{12}6^{4}$) cages. Assuming single full occupancy, the sII unit cell has 24 additional N$_{2}$ or H$_{2}$ molecules in the cases of the N$_{2}$ and H$_{2}$ hydrates, respectively. The case of the THF hydrate is special since the THF molecules only occupy the larger or H cages of the sII structure, with only one THF molecule per cage (single occupancy). According to this, in the solid crystalline structure of this hydrate, we have 8 additional THF molecules. From the thermodynamics point of view, the principal dissociation or three-phase line of the N$_{2}$ and H$_{2}$ hydrates exhibits is a H--L$_{\text{w}}$--V three-phase line at which the hydrate, the aqueous solution of N$_{2}$ or H$_{2}$, and the vapor phases coexist. The main dissociation line of the THF hydrate is an univariant two-phase coexistence line, as we have already mentioned. We recommend the reader the books of Sloan and Koh~\cite{Sloan2008a} and Ripmeester and Alavi~\cite{Ripmeester2022a} for a review of the different structures of hydrates.

Figure ~\ref{figure4} shows the $\overline{q}_{12}-\overline{q}_{3}$ representations of the order parameters for water molecules in aqueous solutions and solid phases of the N$_{2}$, THF, and H$_{2}$ hydrates.       In the case of the N$_{2}$ and H$_{2}$ hydrates, the parameter values are obtained at supercooling conditions (see below). In the case of the THF hydrate, the order parameters have been obtained at coexistence conditions ($500\,\text{bar}$ and $272.5\,\text{K}$). Note that in this case, only two phases coexist. The first conclusion given the results obtained for the three hydrates is that the same $\overline{q}_{12}-\overline{q}_{3}$ representation of the order parameters, valid for the structure sI, can also differentiate liquid-like and hydrate-like water molecules. Fig.~\ref{figure4}a (top panel) shows the representation for the case of the N$_{2}$ hydrate at $500\,\text{bar}$ and $250\,\text{K}$ (supercooling of $35\,\text{K}$). Although it is possible to have a small value of the mislabeling using only a threshold value of $\overline{q}_{12}$, we have opted here for a $\overline{q}_{12}$ and $\overline{q}_{3}$ combination. In this case, the mislabeling is $0.0006\%$. However, in the case of the H$_{2}$ hydrate, it is possible to have a practically constant threshold for the $\overline{q}_{12}$ parameter, $\overline{q}_{12,t}\approx 0.142$ to ensure the complete separation of the clouds associated to the liquid-like (black circles) and the hydrate-like (red circles) water molecules. The corresponding order parameter values, obtained at $1850\,\text{bar}$ and $240\,\text{K}$ (supercooling of $35\,\text{K}$), can be seen in Fig.~\ref{figure4}b (middle panel). In this case, the mislabeling is even lower than in the case of the N$_{2}$ hydrate, below $0.0001\%$. Finally, it is also possible to use the same combination of order parameters to address the identification of liquid-like and hydrate-like water molecules in the case of the THF hydrate. According to Fig.~\ref{figure4}c (bottom panel), the linear combination of the $\overline{q}_{12}$ and $\overline{q}_{3}$ order parameters given a mislabeling of $0.003\%$.

It is interesting to mention that we have explored all possible combinations of the first twenty order parameters, as indicated in the previous section in the case of sI hydrates. In all cases, the best option for describing two separated clouds with the low mislabeling values shown in Fig.~\ref{figure4} correspond to the $\overline{q}_{12}$-$\overline{q}_{3}$ combination. In general, the separation of the two clouds is more effective in the case of the sII hydrates than in the case of the sI ones, especially in the case of the N$_{2}$ and H$_{2}$ molecules. The reason is probably due to two effects. First, the solubility of these molecules in water is very small compared with that of CH$_{4}$ and especially CO$_{2}$. Usually, a high concentration of guest molecules in the liquid water phase produces scattering of the cloud associated with the liquid-like molecules, which results in a larger overlap of the cloud with that of the hydrate-like molecules. Note, for instance, how the cloud (black pluses) of the aqueous solution of THF is wider than the clouds of the aqueous solutions of N$_{2}$ and H$_{2}$. Second, the sII crystallographic structure is characterized by values of the $\overline{q}_{12}$ parameter above $0.2$. This is a higher value compared with that associated to sI hydrate structures (see Figs.~\ref{figure1}-\ref{figure3} and compared with results in Fig.~\ref{figure4}). The combination of these two effects results in a better separation of clouds for hydrate systems that exhibit the sII crystallographic structure.

\subsection{Analysis of order parameters from simulation of interfacial systems}

Order parameters are routinely used for identifying and counting molecules in emerging phases immersed into metastable phases at given thermodynamic conditions. This information is essential to estimate homogeneous nucleation rates and solid-fluid interfacial free energies. Particularly, averaged local bond order parameters of Lechner and Dellago~\cite{Lechner2008a} have been used to estimate the number of water molecules that appear in incipient small clusters of solid phases embedded in a metastable phase in nucleation studies.~\cite{Sanz2013a,Espinosa2014b,Bianco2021a,Lamas2021a,Grabowska2023a} The precise estimation of the size of the critical cluster at supercooling conditions is crucial to predict reliable values of nucleation rates in seeding simulations used in combination with CNT, BF molecular simulations, or when using more rigorous techniques to estimate nucleation rates, such as Forward Flux Sampling,~\cite{Allen2005a} Transition Path Sampling,~\cite{Bolhuis2002a,Lechner2011a,Beckham2011a} Metadynamics,~\cite{Laio2002a,Trudu2006a} and Lattice Mold.~\cite{Espinosa2016e} This is especially important in the case of the Seeding methodology. In this technique, a good choice of parameters is crucial since a small variation in the number of water molecules forming the critical cluster could vary the nucleation rate by several orders of magnitude. This is not the case in the rest of the techniques, in which the nucleation rates are independent of the election of the order parameter.~\cite{Torrie1977a,Auer2001a,Filion2010a,Borrero2007a} The same is true for the Variational Umbrella Seeding technique, a novel method introduced very recently by Gispen \emph{et al.}~\cite{Gispen2024a} for computing nucleation barriers.

\begin{figure}
\includegraphics[width=0.9\columnwidth,angle=0]{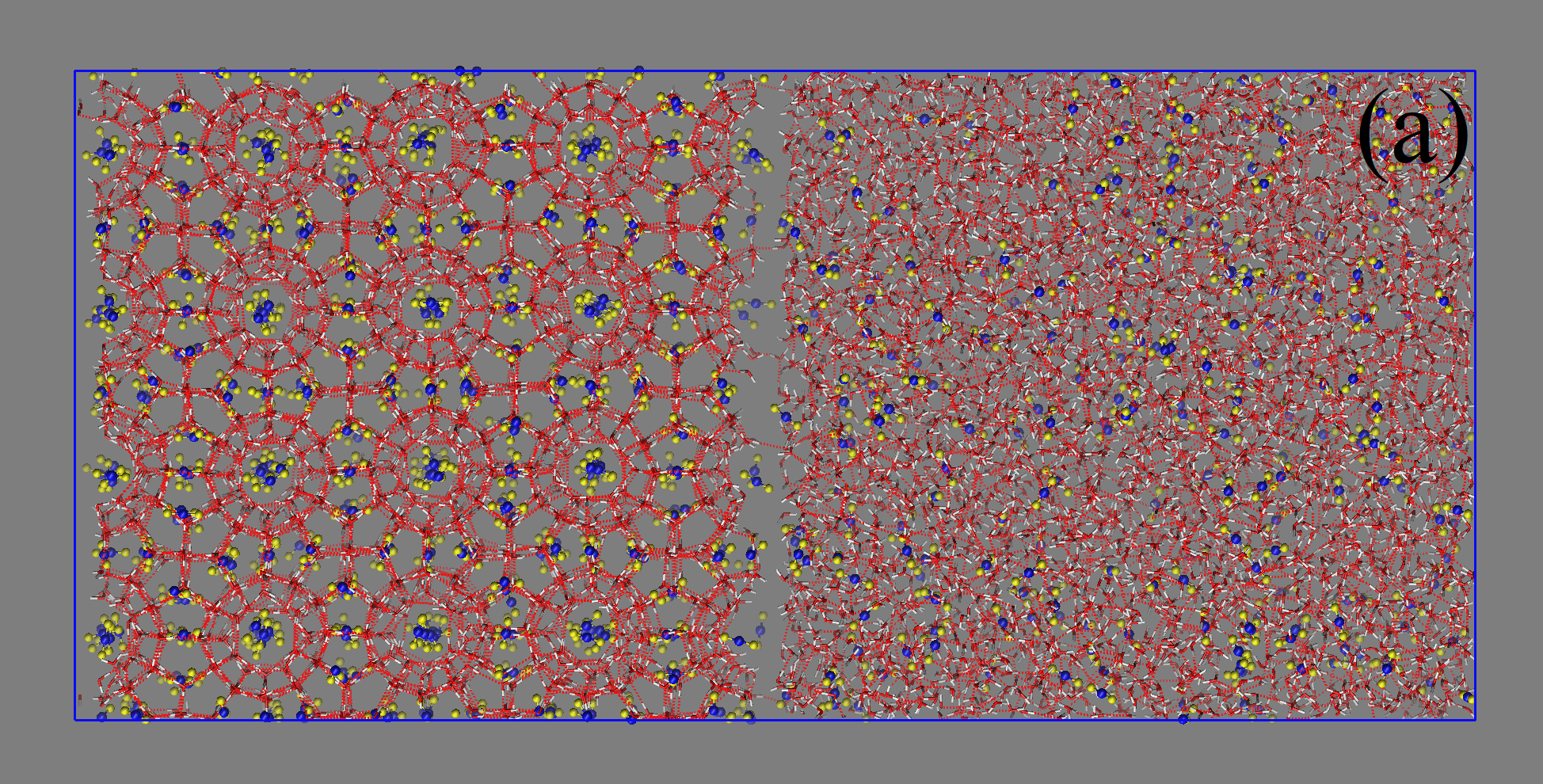}
\includegraphics[width=0.9\columnwidth]{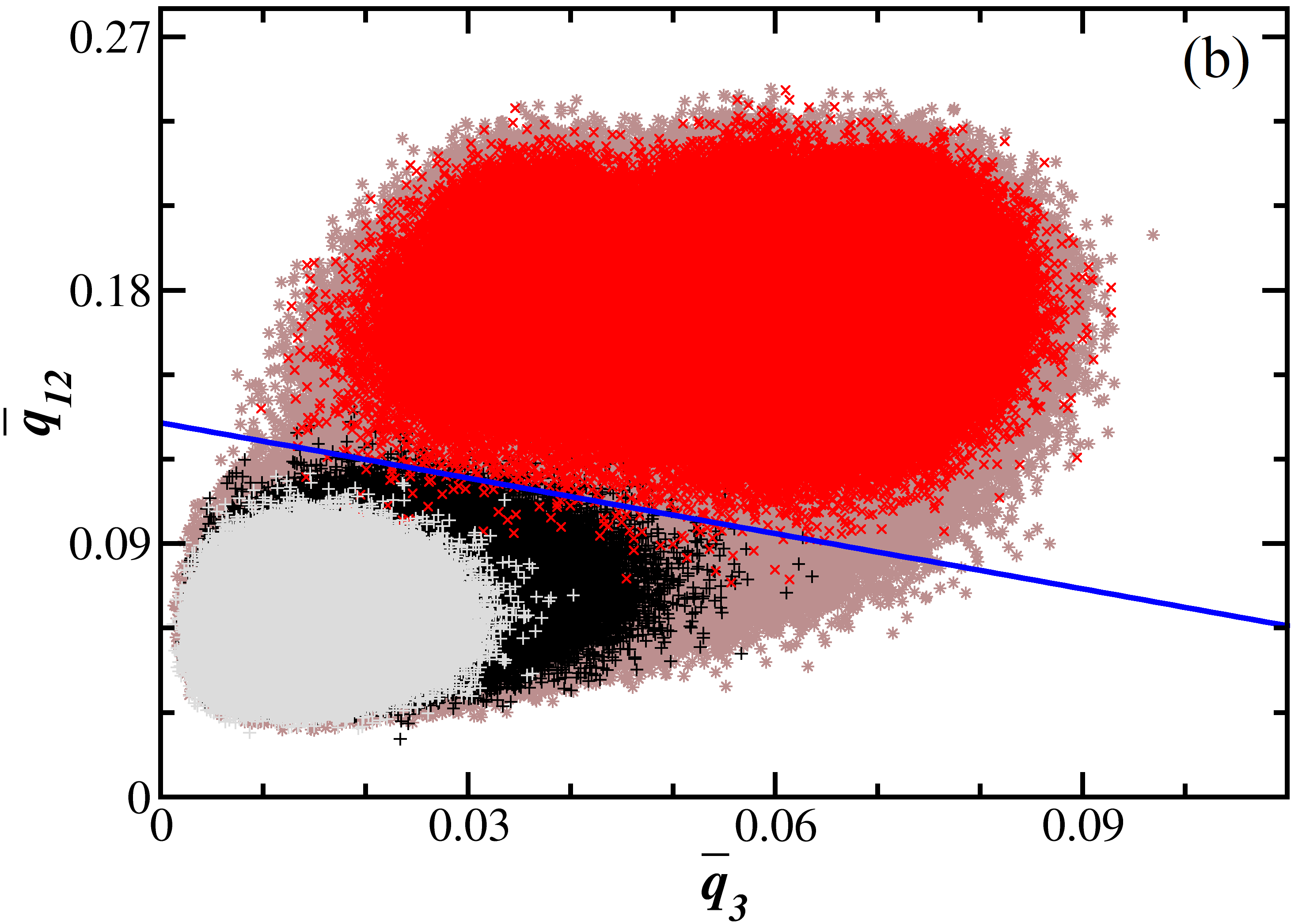}
\caption{(a) Snapshot of the simulation box, at $400\,\text{bar}$ and $287\,\text{K}$, of the CO$_{2}$ hydrate  -- aqueous solution of CO$_2$ two-phase coexistence. Red and white licorice representation corresponds to oxygen and hydrogen atoms of water, respectively, blue and yellow spheres (van der Waals representation) correspond to carbon and oxygen atoms of CO$_{2}$, and dashed red lines to hydrogen bonds between water molecules. (b) Values of $\overline{q}_{12}$ versus $\overline{q}_{3}$ for water molecules at the same thermodynamic conditions in the CO$_2$ hydrate -- aqueous solution of CO$_{2}$ (brown stars), the CO$_{2}$ hydrate one-phase (red crosses), the aqueous solution of CO$_{2}$ one-phase (black pluses), and the pure water (grey pluses) systems.}
\label{figure5}
\end{figure}

The other problem in which an accurate estimation of the number of water molecules forming an emerging solid phase is key is the determination of solid-fluid interfacial free energies using the Mold Integration technique and its extensions.~\cite{Espinosa2014a,Espinosa2016a,Algaba2022b,Zeron2022a,Romero-Guzman2023a} As we have mentioned in the Introduction, this methodology is based on the use of a mold of attractive wells located at the crystallographic positions of the atoms of molecules in the equilibrium solid phase to induce the formation of a thin hydrate slab in the liquid phase at coexistence conditions. One of the key magnitudes in the method is the evaluation of the number of molecules forming the induced solid phase, which is determined using the order parameters of Lechner and Dellago.~\cite{Lechner2008a}

The common point of the problems discussed in the two previous paragraphs is that the systems under consideration exhibit interfaces separating solid and liquid phases. In both cases, bond order parameters are used to analyze if water molecules are hydrate-like or liquid-like from information obtained simulating previously two independent one-phase systems. In this context, the following question arises: is it possible to use these local bond order parameters and the $\overline{q}_{12}$~--~$\overline{q}_{3}$ representation to distinguish with confidence if water molecules are hydrate-like or liquid-like? To answer this question, we concentrate from this point on the use of the $\overline{q}_{12}$~--~$\overline{q}_{3}$ representation of bond order parameters to identify if water molecules are hydrate-like or liquid-like in systems involving CO$_{2}$ hydrates. We consider two different benchmark systems in which hydrate-fluid phases are present. The first example corresponds to a standard coexistence between a hydrate and an aqueous solution phases via a planar interface and the second one to the application of the Mold Integration technique to determine hydrate-water interfacial free energies.

In the first example, we consider a system formed by an aqueous solution of CO$_{2}$ in contact with a CO$_{2}$ hydrate phase via a planar interface at $287\,\text{K}$ and $400\,\text{bar}$. The system involves $2944$ and $510$ water and CO$_{2}$ molecules forming the hydrate phase. This corresponds to a hydrate phase formed by $4\times4\times4$ unit cells of the CO$_{2}$ hydrate (the unit cell is replicated four times along each direction of the space). The aqueous solution is formed from $4000$ and $240$ water and CO$_{2}$ molecules, respectively. This corresponds to a molar fraction $x_{\text{CO}_{2}}\approx0.057$, the equilibrium concentration of CO$_{2}$ in water in equilibrium with the hydrate phase at $287\,\text{K}$ and $400\,\text{bar}$.~\cite{Miguez2015a,Algaba2022b,Zeron2022a,Romero-Guzman2023a,Algaba2023a} Fig.~\ref{figure5}a shows a snapshot of the final configuration of the system in which the CO$_{2}$ hydrate phase (left side) coexists with an aqueous solution of CO$_{2}$ (right side) via a planar interface. Due to 
the periodic boundary conditions used during the simulation, the system exhibits two hydrate--aqueous solution or H--L$_{\text{w}}$ interfaces (L$_{\text{w}}$--H--L$_{\text{w}}$ system).

We use an equilibration period of $2\,\text{ns}$ and $50\,\text{ns}$ for the production stage. We determine the $\overline{q}_{3}$ and $\overline{q}_{12}$ local order parameter values for water molecules in the H--L$_{\text{w}}$ interfacial simulation box. Fig.~\ref{figure5}b shows the 
$\overline{q}_{3}$--$\overline{q}_{12}$  representation of the order parameters in the simulation box at $287\,\text{K}$ and $400\,\text{bar}$. The $\overline{q}_{3}$ and $\overline{q}_{12}$ values are calculated during the production time of the simulation (brown stars). For comparison reasons, we have also included the values of the $\overline{q}_{3}$ and $\overline{q}_{12}$ order parameters obtained simulating the hydrate one-phase (red crosses) and the aqueous one-phase (black pluses) separately at the same thermodynamic conditions of coexistence. Note again that this representation is the same as in Fig.~\ref{figure1}c. In addition to that, we also represent the values of the parameters simulating a pure water one-phase system at the same conditions (grey pluses). Note that the cloud obtained simulating the pure water one-phase is narrower than the cloud corresponding to the aqueous solution phase of CO$_{2}$. The absence of CO$_{2}$ in the liquid phase alters the local environment of water molecules, producing larger values of $\overline{q}_{3}$ and $\overline{q}_{12}$  and making less distinguishable hydrate-like and liquid-like water molecules. Nevertheless, values obtained simulating hydrate and liquid one-phase systems produce two different and well-separated clouds. As can be seen in Fig.~\ref{figure5}b, the parameter values obtained from the simulation of the H--L$_{\text{w}}$ interfacial box are distributed in a slightly wider region (marron stars) than those corresponding to the union of the clouds associated with the hydrate-like (red crosses) and liquid-like (black pluses) water molecules. Clearly, it is difficult to distinguish hydrate-like and water-like water molecules in the regions outside the red and black clouds but this is expected from the use of local order parameters when interfaces are present. Particularly, all the points are bounded in the region defined by $\overline{q}_{3}\lesssim0.1$ and $\overline{q}_{12}\lesssim0.27$.

We now consider the second benchmark system, a fluid mixture of CO$_{2}$ and water at $287\,\text{K}$ and $400\,\text{bar}$. These thermodynamic conditions correspond to a thermodynamic state at the dissociation or three-phase line of the CO$_{2}$ hydrate at which a hydrate phase, an aqueous solution of CO$_{2}$ with the equilibrium composition, and a CO$_{2}$-rich liquid phase coexist. According to the Mold Integration Guest methodology,~\cite{Algaba2022b,Zeron2022a,Romero-Guzman2023a} we simulate a system formed from two liquid phases, a water-rich liquid phase and a CO$_{2}$-rich liquid phase, separated by a planar interface. We perform two different simulations in the $NP_{z}\mathcal{A}T$ ensemble at $287\,\text{K}$ and $400\,\text{bar}$. In the first one, the interaction sites of the mold are switched off. In this case, no hydrate phase exists during the simulation and only the two liquid phases previously described are present during the simulations. In the second one, we perform simulations with the interaction sites of the mold switched on. In this case, and according to the Mold Integration technique, a hydrate phase is induced around the crystallographic positions of the mold.~\cite{Algaba2022b,Zeron2022a,Romero-Guzman2023a} In both cases, we use an equilibration period of $2\,\text{ns}$ and $50\,\text{ns}$ for the production stage. Following a similar approach to that used for the first benchmark system, during this last period, all trajectories are analyzed using the $\overline{q}_{12}$~--~$\overline{q}_{3}$ representation of the order parameters.

\begin{figure}
\includegraphics[width=0.9\columnwidth]{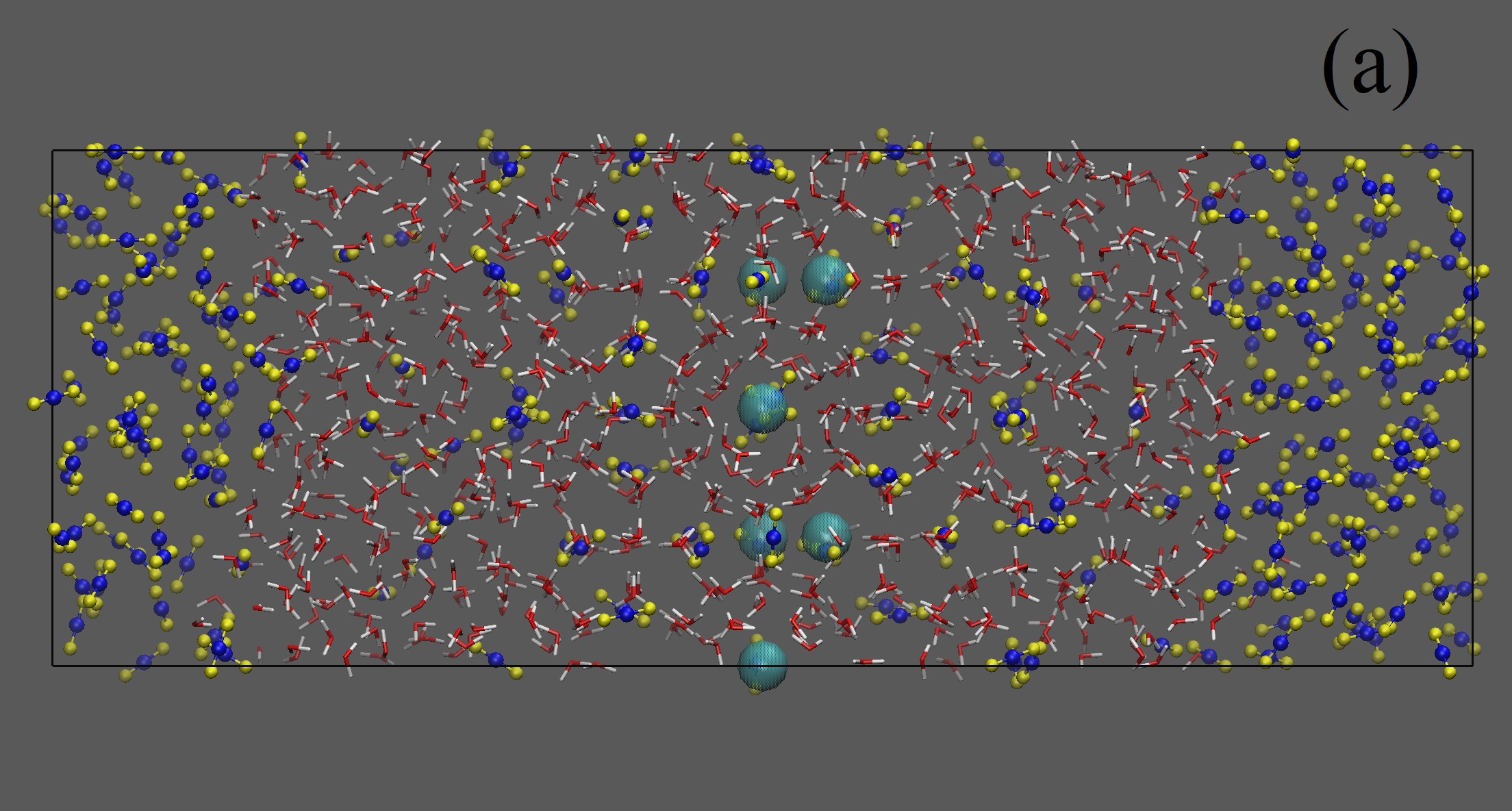}
\includegraphics[width=0.9\columnwidth]{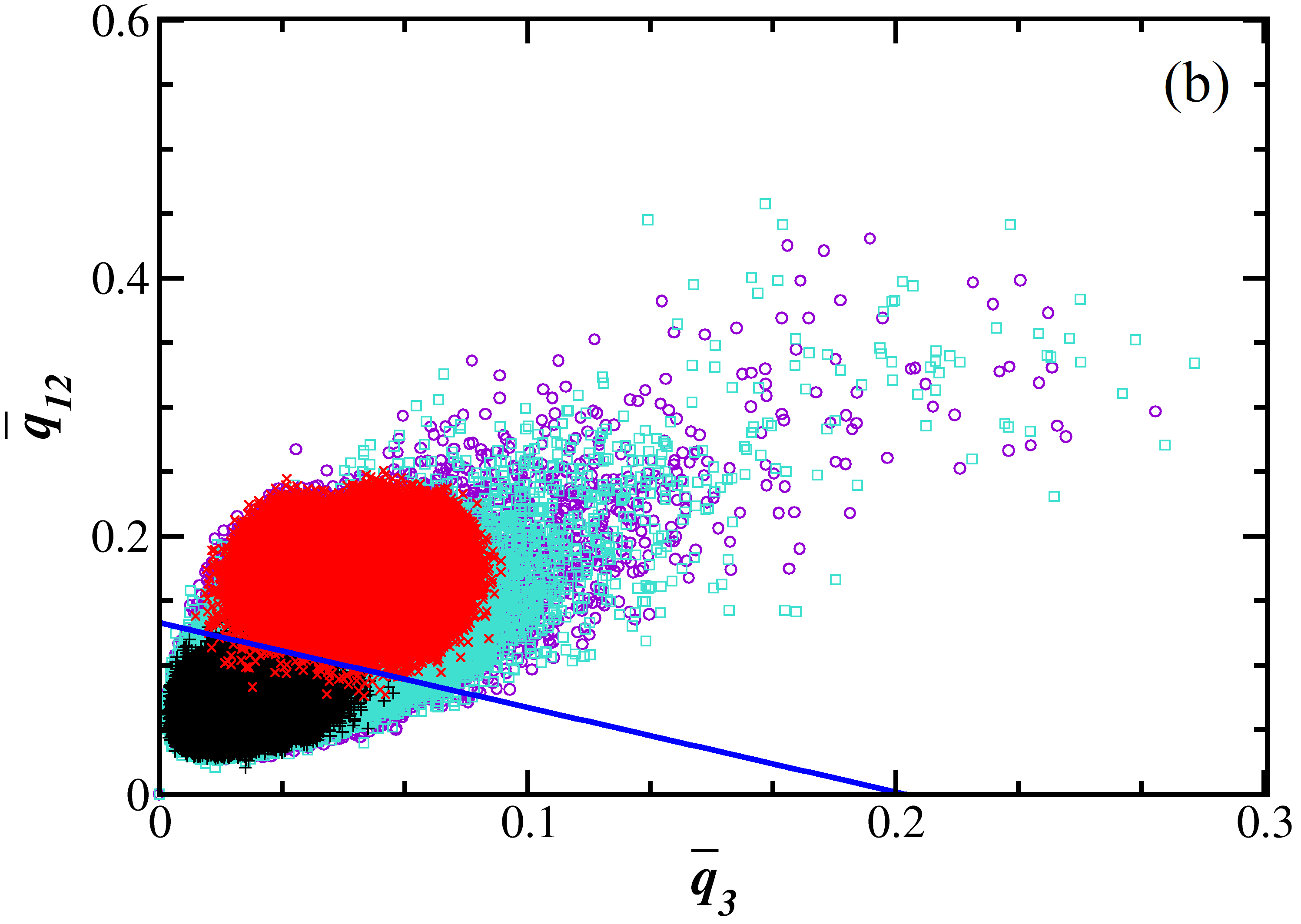}
\caption{(a) Snapshot of the L$_{\text{CO}_{2}}$--L$_{\text{w}}$--H--L$_{\text{w}}$--L$_{\text{CO}_{2}}$ simulation box showing the crystallization of the CO$_{2}$ hydrate phase (center) from the CO$_{2}$ + water two-phase coexistence at $400\,\text{bar}$ and $287\,\text{K}$. Red and white licorice representation corresponds to oxygen and hydrogen atoms of water, respectively, yellow and blue spheres (Van der Waals representation) correspond to carbon and oxygen atoms of CO$_{2}$, and cyan spheres (van der Waals representation) correspond to the mold attractive sites. (b) Values of $\overline{q}_{12}$ versus $\overline{q}_{3}$ for water molecules, at the same thermodynamic conditions, in the system, when the interaction sites of the mold are switched on (violet circles) and switched off (cyan squares), in the CO$_2$ hydrate one-phase (red crosses), and the aqueous solution of CO$_{2}$ one-phase (black pluses).}
\label{figure6}
\end{figure}

Figure~\ref{figure6}a shows a snapshot of a representative configuration of the system showing the aqueous solution of CO$_{2}$ in the center of the simulation box, the CO$_{2}$-rich liquid phase surrounding the aqueous phase at the borders of the box, and some interactions sites in the aqueous phase at the equilibrium positions in one of the principal planes of the crystallographic sI structure of the CO$_{2}$ hydrate. The configuration shown in Figure~\ref{figure6}a corresponds to the final configuration of a simulation in which the interaction sites of the mold are switched on. According to the Mold Integration method, during simulation, a hydrate phase appears in the center of the simulation box, around the interaction sites of the mold. The final configuration contains a hydrate phase in the center of the simulation box, an aqueous solution of CO$_{2}$ surrounding the hydrate phase, and the CO$_{2}$-rich liquid phase at the border of the simulation box. Due to the use of periodic boundary conditions in the simulation, we have an L$_{\text{CO}_{2}}$--L$_{\text{w}}$--H--L$_{\text{w}}$--L$_{\text{CO}_{2}}$ systems, with two 
L$_{\text{CO}_{2}}$--L$_{\text{w}}$ interfaces and two L$_{\text{w}}$--H interfaces.

In Figure~\ref{figure6}b, we present the $\overline{q}_{12}$~--~$\overline{q}_{3}$ representation of the local bond order parameters for water molecules in the L$_{\text{CO}_{2}}$--L$_{\text{w}}$--H--L$_{\text{w}}$--L$_{\text{CO}_{2}}$ (interfacial simulation box) at $400\,\text{bar}$ and $287\,\text{K}$. The values of 
$\overline{q}_{3}$ and $\overline{q}_{12}$ shown are calculated during the simulation when using the Mold Integration technique with the interaction sites of the mold switched on (open violet circles) but also when they are switched off (open cyan squares). It is important to recall that when the sites are switched off there is no solid phase in the system and only two liquid-liquid (L$_{\text{CO}_{2}}$--L$_{\text{w}}$) interfaces exist. For comparison reasons, we have also included, as in the case of the first benchmark system, the $\overline{q}_{3}$ and $\overline{q}_{12}$ order parameter values obtained simulating the hydrate one-phase (red cross) and the aqueous one-phase (black pluses) separately at the same thermodynamic conditions of coexistence. This representation has already been presented in Figs.~\ref{figure1}c and \ref{figure5}b. As can be seen, the order parameter values obtained simulating independently the hydrate and aqueous one-phase systems are distributed in two different but well-separated clouds that allow distinguishing between hydrate-like (red crosses) and liquid-like water molecules (black pluses). They are separated by the linear combination of the $\overline{q}_{3}$ and $\overline{q}_{12}$ parameters determined by a straight blue line representing 
the linear relationship discussed previously in Fig.~\ref{figure1}c.

However, the distribution of the $\overline{q}_{3}$ and $\overline{q}_{12}$ parameter values obtained from the interfacial simulation box with the interaction sites of the mold switched on (open violet circles) is quite different. The parameter values are distributed in a wider region than those corresponding to the clouds associated with hydrate-like (red crosses) and liquid-like (black pluses) water molecules. The plot indicates that there are not two different and separated clouds. In other words, it is not possible to distinguish between both types of water molecules. In addition, there exists a dispersion of parameter values towards the region identified as the hydrate phase (hydrate-like water molecules). According to the standard interpretation of the $\overline{q}_{12}$~--~$\overline{q}_{3}$ performed in Section IV.A, each violet square located above the straight blue line is associated with a $(\overline{q}_{3},\overline{q}_{12})$ combination labeled as a hydrate-like water molecule. Can these dispersed points, above the blue straight line, identified as solid-like water molecules?

To confirm or deny this question, we have analyzed the same system (interfacial simulation box) but with the interaction sites of the mold switched off. According to the Mold Integration technique, there is no induction of the hydrate phase and only two liquid-liquid interfaces exist (L$_{\text{CO}_{2}}$--L$_{\text{w}}$). The order parameter values obtained when the mold is switched off are represented using open cyan squares. As can be seen, the $\overline{q}_{12}$~--~$\overline{q}_{3}$ representation order parameters for water molecules indicate the same qualitative behavior obtained when the mold is switched on (open violet circles): a large dispersion of points above the blue straight line associated with hydrate-like water molecules. However, these water molecules are wrongly identified with solid-like water molecules since with the sites of the mold switched off, there is no solid phase. Particularly, we now have points with $\overline{q}_{3}\gtrsim0.1$ nor with $\overline{q}_{12}\gtrsim0.27$. This is contrary to what happens in Fig.~\ref{figure5}b where there does not exist a dispersion of points ($\overline{q}_{3}\lesssim0.1$ and $\overline{q}_{12}\lesssim0.27$). 

It is interesting to mention that we have also simulated an aqueous solution of CO$_{2}$ in contact with vacuum. We have also simulated a pure water phase in contact with vacuum. In both cases, we have obtained the same qualitative behavior of the $\overline{q}_{3}$ and $\overline{q}_{12}$ order parameters shown in Fig.~\ref{figure6}b (not presented here). Why the $\overline{q}_{3}$--$\overline{q}_{12}$  representation of the local order parameters for water molecules in Figs.~\ref{figure5}b (L$_{\text{w}}$--H--L$_{\text{w}}$ system) and \ref{figure6}b (L$_{\text{CO}_{2}}$--L$_{\text{w}}$--H--L$_{\text{w}}$--L$_{\text{CO}_{2}}$ system) are so different? What is going on? We think that the presence of interfaces in the simulation box significantly alters the local environment of water molecules and the criterion used to identify water molecules as solid-like or liquid-like fails when using the local order parameters obtained using one-phase simulations in conditions at which interfaces exist. But is it only due to the presence of any interface or some particular interfaces?

\begin{figure}
\includegraphics[width=0.9\columnwidth]{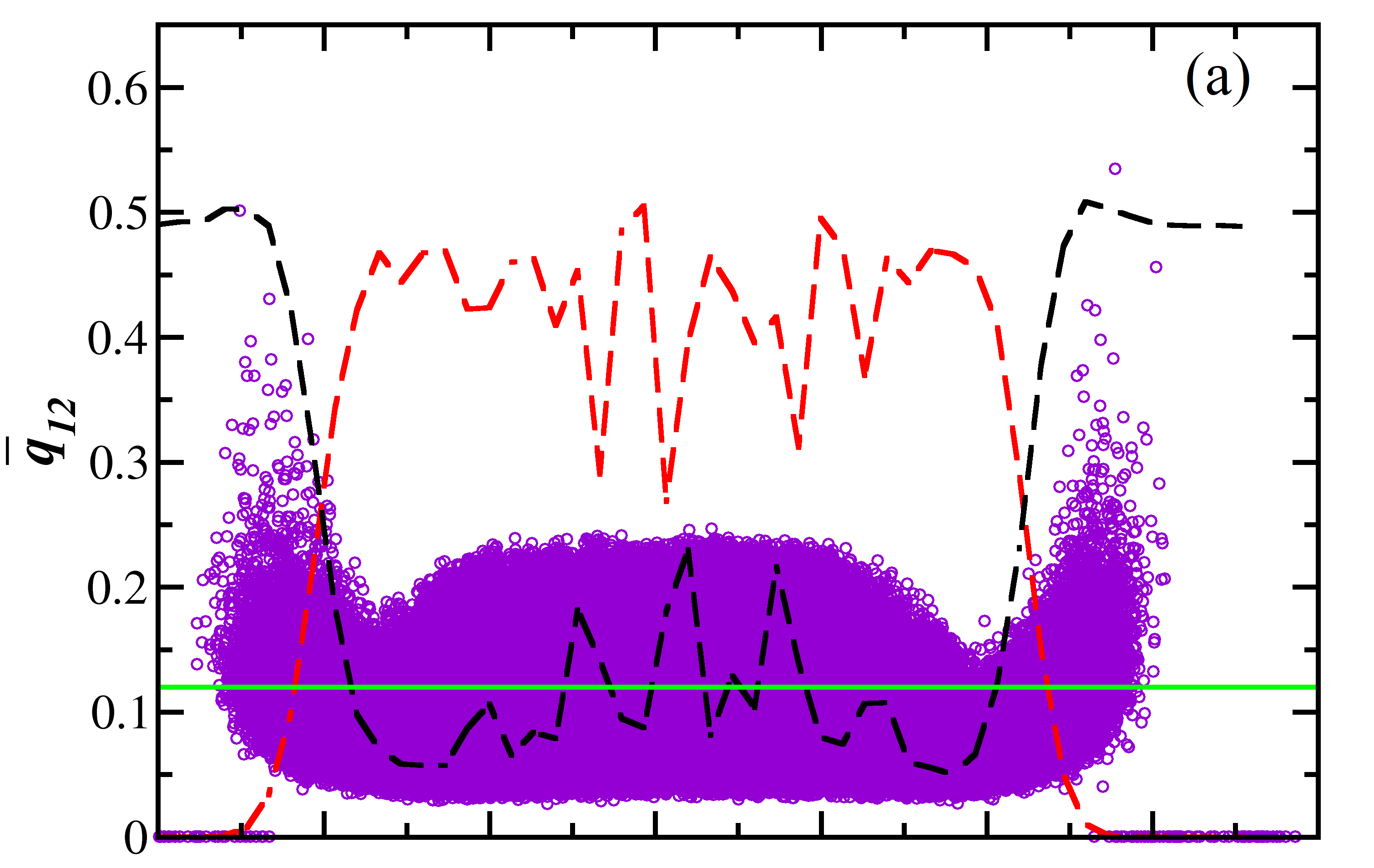}
\includegraphics[width=0.9\columnwidth]{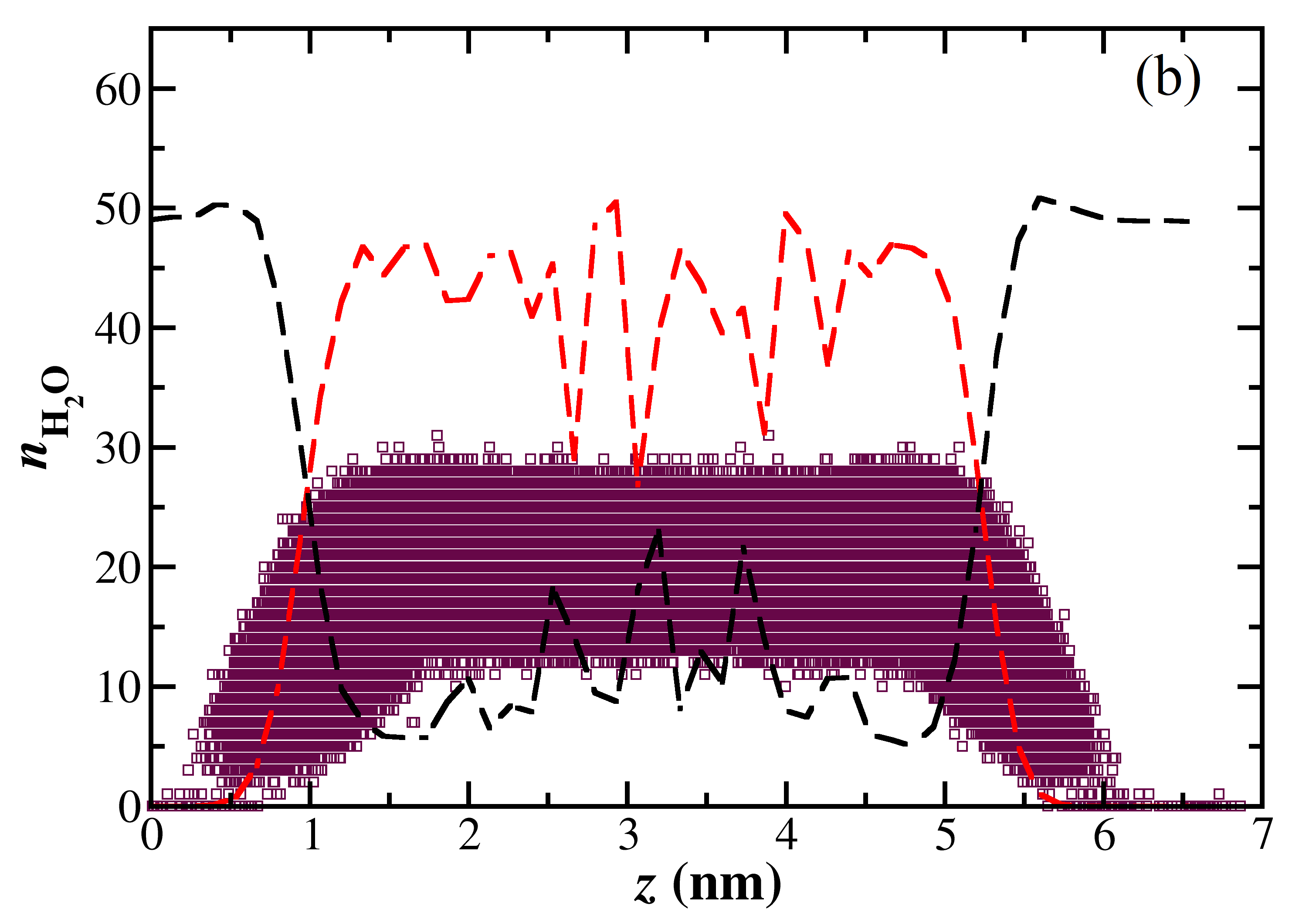}
\caption{(a) Values of $\overline{q}_{12}$ (violet circles) and (b) number of water neighbors (violet circles), $n_{\text{H}_2\text{O}}$, as functions of $z$ along the L$_{\text{CO}_{2}}$--L$_{\text{w}}$--H--L$_{\text{w}}$--L$_{\text{CO}_{2}}$ simulation box at $400\,\text{bar}$ and $287\,\text{K}$. Density profiles of water (red dashed curve) and CO$_{2}$ (black dashed curve) along the $z$ and the $\overline{q}_{12,t}$ threshold value (horizontal green line) are also shown.}
\label{figure7}
\end{figure}

To clarify the results obtained in Fig.~\ref{figure6}b and understand why some water molecules are considered hydrate-like (those with $\overline{q}_{3}$ and $\overline{q}_{12}$ values above the corresponding threshold), we have considered the $\overline{q}_{12}$ order parameter values associated to the water molecules along the $z$-axis, perpendicular to the interfaces exhibited by the L$_{\text{CO}_{2}}$--L$_{\text{w}}$--H--L$_{\text{w}}$--L$_{\text{CO}_{2}}$ system. Note that this system exhibits four interfaces, two L$_{\text{w}}$--L$_{\text{CO}_{2}}$ and two H--L$_{\text{w}}$ interfaces. Note that it is also possible to use the $\overline{q}_{3}$ order parameter for this discussion. Fig.~\ref{figure7}a shows the $\overline{q}_{12}$ order parameter, as a function of the $z$ coordinate of the corresponding water molecule. To clarify the discussion, we have also included the scaled density profiles of water (dashed red curve) and CO$_{2}$ molecules (dashed black curve) along the interfaces. Note that the simulation box corresponds to the system in which the Mold Integration technique is used to determine the hydrate--water interfacial free energy. According to this, the interaction sites of the mold are switched on and a hydrate phase is induced in the center of the simulation box (see the peaks of the density profiles of water and CO$_{2}$ molecules showing the crystallographic equilibrium positions of the molecules in the sI hydrate structure at $1.5\lesssim z\lesssim 5\,\text{nm}$). In addition, since water is not miscible in CO$_{2}$ at these conditions, the density profiles of water at the borders of the simulation box are equal to zero, as expected, and the density of CO$_{2}$ approaches asymptotically to the bulk values at $287\,\text{K}$ and $400\,\text{bar}$. 

Figure~\ref{figure6}b indicates that the threshold value of $\overline{q}_{12}$ that allows to distinguish between hydrate- and liquid-like water molecules is $\overline{q}_{12,t}\approx 0.11$. According to this, water molecules with $\overline{q}_{12}>\overline{q}_{12,t}$ are hydrate-like and those with $\overline{q}_{12}<\overline{q}_{12,t}$ are liquid-like. This is indicated in Fig.~\ref{figure7}a as a horizontal green line. As can be seen, in the region at which the system exhibits the CO$_{2}$ hydrate phase, the $\overline{q}_{12}$ order parameter values are clearly below the threshold value ($\overline{q}_{12}\lesssim 0.25$). In other words, a large number of water molecules are identified as hydrate-like molecules. However, note that there are also water molecules in the same region identified as liquid-like as we observe  $\overline{q}_{12}$ values below  $\overline{q}_{12,t}$. This is an expected and not avoidable effect. These $\overline{q}_{12}$ values, associated with liquid-like molecules, correspond to water molecules detected as liquid-like before the hydrate grew or even when the hydrate melted during the simulation since the $\overline{q}_{12}$ values are collected from the beginning of the simulation when the interaction sites of the mold are still switched off and the hydrate is not formed.

The behavior of the $\overline{q}_{12}$ local order parameter at the L$_{\text{w}}$--L$_{\text{CO}_{2}}$ interfacial region and in the CO$_{2}$-rich liquid phase, for $z\lesssim 1.5\,\text{nm}$ and $z\gtrsim 5\,\text{nm}$, deserves a more detailed analysis. As can be seen, there is a large dispersion of points with high $\overline{q}_{12}$ values ($\overline{q}_{12}>\overline{q}_{12,t}$). Hence, these points are associated with water molecules identified as hydrate-like molecules. However, these water molecules are not in the hydrate phase but are located at the 
L$_{\text{w}}$--L$_{\text{CO}_{2}}$ interface according to the density profiles of water and CO$_{2}$. It is important to recall that we find the $\overline{q}_{12}$ diverge as $z$ values are closed to the L$_{\text{w}}$--L$_{\text{CO}_{2}}$ interfacial regions. We have set to zero these $\overline{q}_{12}$ in the figure to have a clear and neat representation of the rest of the values.

It is also possible to calculate the number of neighbors of each water molecule, $n_{\text{H}_{2}\text{O}}$, and correlate this information with the position of the molecule in the simulation box and the associated $\overline{q}_{12}$ local order parameter value. Note that in this work, $n_{\text{H}_{2}\text{O}}$ is the number of molecules/neighbors inside a sphere of radius $\sim 0.55\,\text{nm}$ centered at a given water molecule. This is the cutoff value used to identify neighbors using the Lechner and Dellago local order parameters~\cite{Lechner2008a} (see Section II for further details). Fig.~\ref{figure7}b shows $n_{\text{H}_{2}\text{O}}$ as a function of $z$. As can be seen, in the center of the simulation box ($1.5\lesssim z\lesssim 5\,\text{nm}$), the number of neighbors of water molecules is approximately $10-30$. This result agrees with the behavior of the $\overline{q}_{12}$ order parameter of the same water molecules located in the same region of the simulation box. However, this is not the case for the water molecules located in the L$_{\text{w}}$--L$_{\text{CO}_{2}}$ interfaces, closed to the CO$_{2}$-rich liquid phase ($1.5\lesssim z\lesssim 5\,\text{nm}$). The $\overline{q}_{12}$ values associated with these water molecules are above the threshold value $\overline{q}_{12,t}$, i.e., they are identified as hydrate-like molecules. However, as $z\rightarrow 0$ (left border of the simulation box) and $z\rightarrow 7\,\text{nm}$ (right border of the simulation box), the number of neighbors of water molecules decreases. This is expected behavior since each water molecule has no neighbors in the CO$_{2}$-rich liquid phase. This corroborates that the $\overline{q}_{12}$ local order parameter takes unrealistically high values for water molecules not surrounded by other water molecules, as it shows Fig.~\ref{figure7}b, and identifying them wrongly as hydrate-like.

\begin{figure}
\includegraphics[width=0.9\columnwidth]{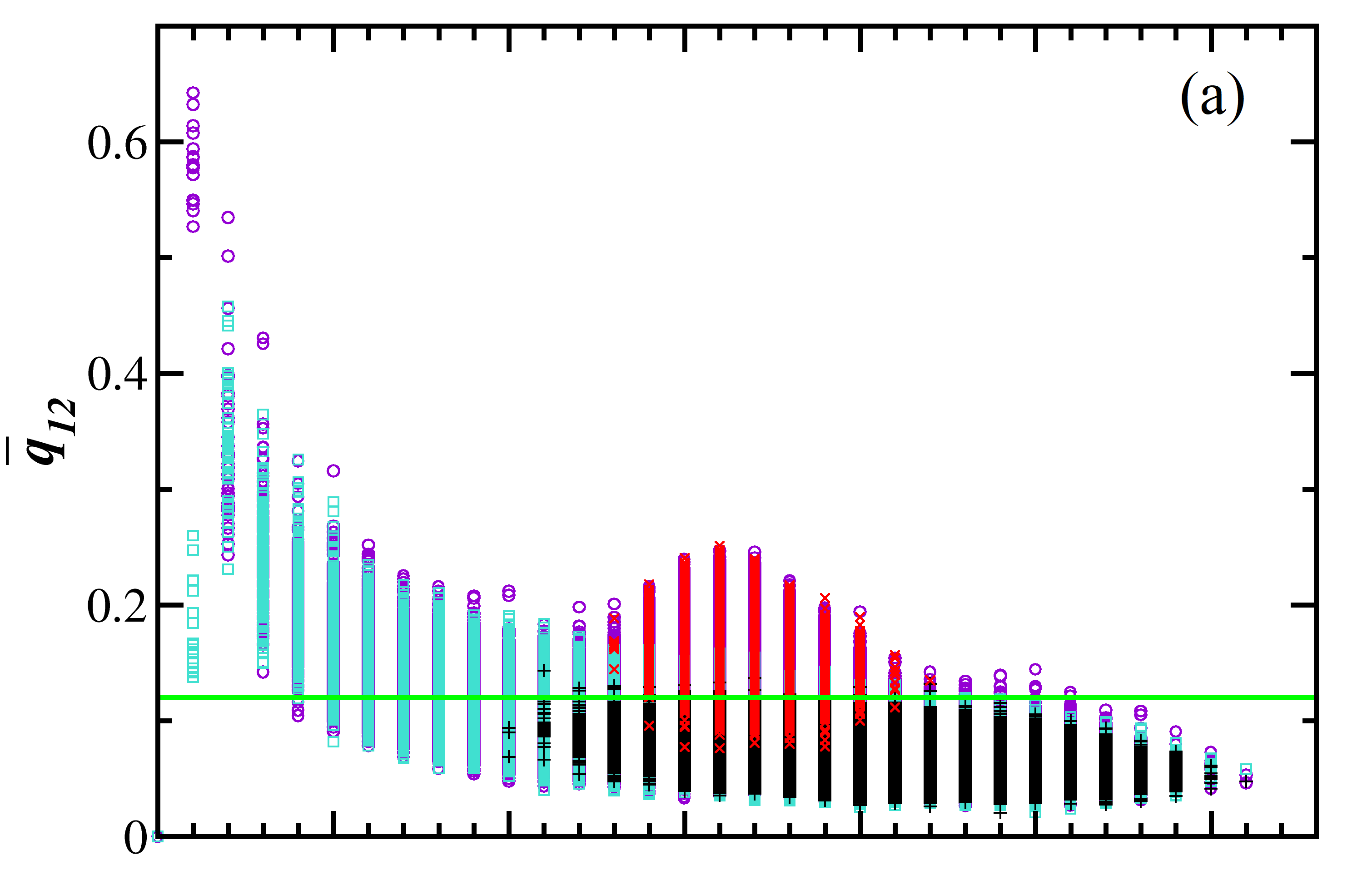}
\includegraphics[width=0.9\columnwidth]{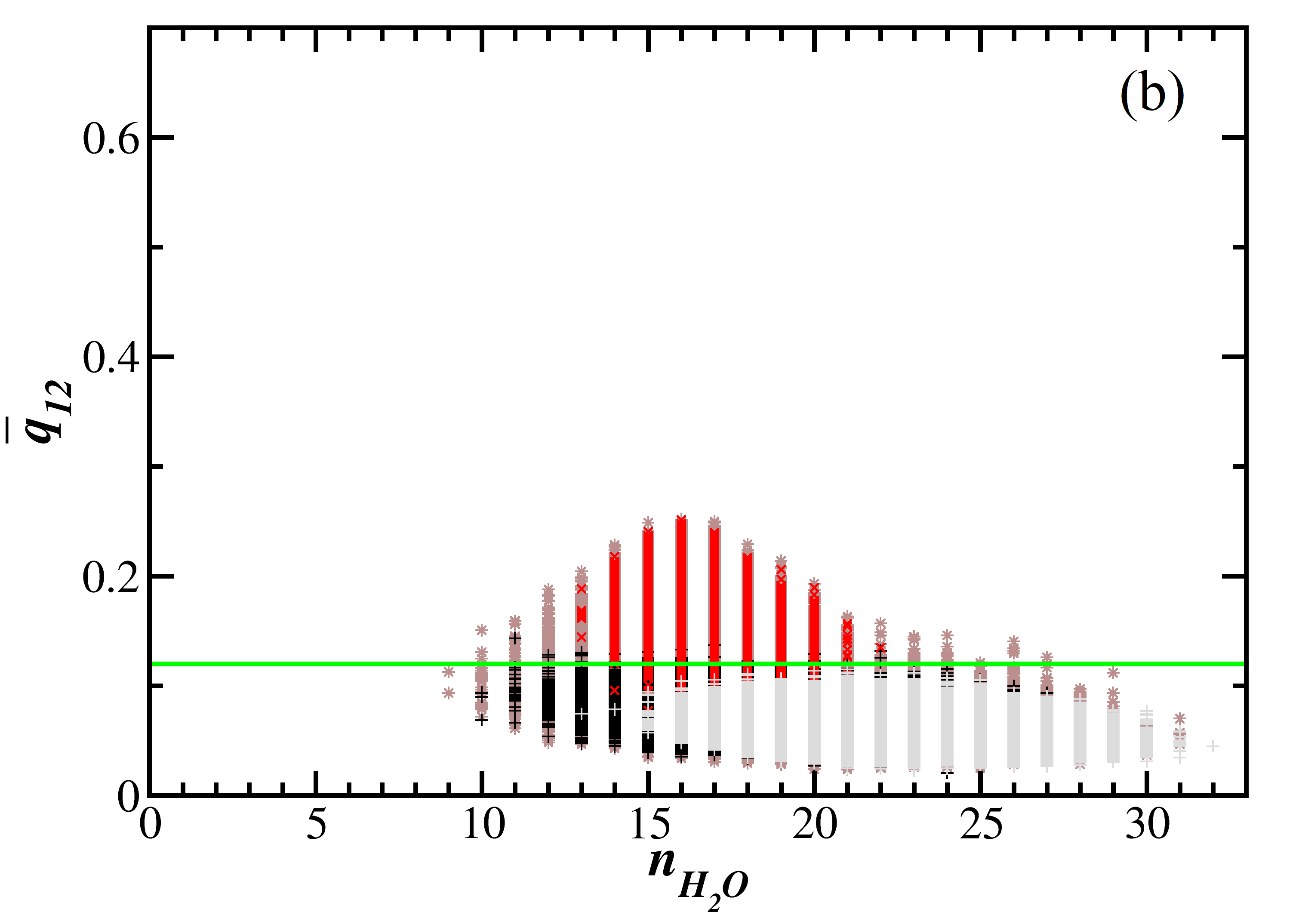}
\caption{(a) Values of $\overline{q}_{12}$, as a function of $n_{\text{H}_2\text{O}}$, for water molecules in the L$_{\text{CO}_{2}}$--L$_{\text{w}}$--H--L$_{\text{w}}$--L$_{\text{CO}_{2}}$ simulation box at $400\,\text{bar}$ and $287\,\text{K}$ when the interaction sites of the mold are switched on (violet circles) and switched off (cyan squares). (b) Values of $\overline{q}_{12}$, as a function of $n_{\text{H}_2\text{O}}$, for water molecules at the same thermodynamic conditions in the CO$_{2}$ hydrate - aqueous solution of CO$_{2}$ two-phases coexistence (brown stars) and pure water one-phase (grey pluses) systems. We have also included in (a) and (b) the $\overline{q}_{12}$ values for water molecules in the CO$_{2}$ hydrate one-phase (red crosses) and the CO$_2$ aqueous solution one-phase (black pluses). The $\overline{q}_{12,t}$ threshold value is also shown as a horizontal green line in both panels.}
\label{figure8}
\end{figure}

Finally, Fig.~\ref{figure8}a shows the $\overline{q}_{12}$ local order parameter as a function of the number of neighbors for each water molecule, $n_{\text{H}_{2}\text{O}}$. We have also included the threshold value of $\overline{q}_{12}$ that allows us to distinguish between hydrate-like and liquid-like water molecules (horizontal green line). As in Fig.~\ref{figure6}, the values of $\overline{q}_{12}$ have been calculated during the simulation study using the Mold Integration technique with the interaction sites switched on (open violet circles) but also when they are switched off (open cyan squares). In the first case, we simulate a L$_{\text{CO}_{2}}$--L$_{\text{w}}$--H--L$_{\text{w}}$--L$_{\text{CO}_{2}}$ system and in the second case the simulation results correspond to a L$_{\text{CO}_{2}}$--L$_{\text{w}}$--L$_{\text{CO}_{2}}$ system. We have also included the  $\overline{q}_{12}$ values obtained simulating the CO$_{2}$ hydrate one-phase (red crosses),  and the aqueous solution of CO$_{2}$ one-phase (black pluses). As can be seen, the $\overline{q}_{12}$ values increase as the number of neighbors of water molecules decreases in the cases in which the system exhibits a L$_{\text{CO}_{2}}$--L$_{\text{w}}$ interface. Note that $\overline{q}_{12}$ diverges asymptotically to large values as $n_{\text{H}_{2}\text{O}}\rightarrow 0$. This result confirms the conclusions from Figs.~\ref{figure6} and \ref{figure7}. Since low $n_{\text{H}_{2}\text{O}}$ values are associated to water molecules in fluid phases, the high $\overline{q}_{12}$ values identify wrongly water molecules as hydrate-like.

Fig.~\ref{figure8}b shows the $\overline{q}_{12}$ local order parameter as a function of $n_{\text{H}_{2}\text{O}}$ of systems in which no L$_{\text{CO}_{2}}$--L$_{\text{w}}$ interfaces exist: the first benchmark system with a hydrate-fluid interface (Fig.~\ref{figure5}), an aqueous solution of CO$_{2}$ one-phase, a CO$_{2}$ hydrate one-phase, and a pure water one-phase system. According to the conclusion of the previous paragraph, since no L$_{\text{CO}_{2}}$--L$_{\text{w}}$ interfaces exist, the $\overline{q}_{12}$ local order parameter should exhibit the expected values without the divergent values shown in Fig.~\ref{figure8}a. As can be seen, the number of neighbors of water molecules is between $8$ and $33$ in all cases. In the case of pure water (grey pluses) and the aqueous solution of CO$_{2}$ (black pluses) one-phase systems, $\overline{q}_{12}<\overline{q}_{12,t}$, indicating that there are not hydrate-like water molecules in the systems. However, in the case of the CO$_{2}$ hydrate (red crosses) one-phase system $\overline{q}_{12}>\overline{q}_{12,t}$ for most of the molecules, as expected. The case of the hydrate-fluid (brown stars) one-phase system shows $\overline{q}_{12}$ above and below the threshold value $\overline{q}_{12,t}$ depending on if the water molecule analyzed is hydrate-like and liquid-like, respectively.

To summary, and according to the results presented in Figs.~\ref{figure7} and \ref{figure8}, a good receipt to solve the problem of non-physical high values of the $\overline{q}_{12}$ local order parameter when dealing with systems in which L$_{\text{CO}_{2}}$--L$_{\text{w}}$ interfaces exist is to consider a water molecule as hydrate-like one should use the criterium of the threshold values of $\overline{q}_{12}$ (either $\overline{q}_{3}$ or the $\overline{q}_{3}$ -- $\overline{q}_{12}$ combination), i.e., $\overline{q}_{12}>\overline{q}_{12,t}$, with a number of neighbors equal or greater than $7-8$ ($n_{\text{H}_{2}\text{O}}>7-8$). This ensures that the identification of water molecules as hydrate-like is not artificially overestimated due to the existence of non-physical high $\overline{q}_{12}$ values. The same conclusion is obtained for the analysis of the $\overline{q}_{3}$ values.


\section{Conclusions}

We have revisited and extended the averaged local bond order parameters based on spherical harmonics proposed by Lechner and Dellago~\cite{Lechner2008a} to determine common sI and sII hydrate structures from molecular simulation, including CO$_{2}$, CH$_{4}$, N$_{2}$, H$_{2}$, and THF hydrates. To this end, we have used several molecular models to simulate water, CO$_{2}$, CH$_{4}$, N$_{2}$, H$_{2}$, and THF. Particularly, we have the TIP4P/Ice for water molecules,~\cite{Abascal2005a} the TraPPE models for CO$_{2}$ and N$_{2}$,~\cite{Potoff2001a} a rigid version of the TraPPE force field for THF molecules, a single spherical Lennard-Jones site for CH$_{4}$,~\cite{Guillot1993a,Paschek2004a} and a modified Sivera-Goldmann potential for H$_{2}$.~\cite{Michalis2022a,Silvera1978a,Alavi2005a}

We first analyze the behavior of order parameters obtained from separated simulations of one-phase systems involving aqueous solutions and hydrate phases. We use the standard and most used in the literature local bond order parameters, $\overline{q}_{4}$ and $\overline{q}_{6}$, indicated for distinguishing solid- and liquid-like molecules in model systems (HS and LJ) and realistic force fields for water. Our results suggest that these order parameters are not effective in differentiating between hydrate-like and liquid-like water molecules in hydrate systems.

We have searched combinations of averaged bond parameters that optimize the separation of clouds associated with hydrate- and liquid-like water molecules. To this end, we have analyzed the values of the averaged bond order parameter from $\overline{q}_{1}$ to $\overline{q}_{20}$. We have found that the best combination of all the possible considered in this work is that in which one of the averaged bond parameters is $\overline{q}_{12}$. Particularly, the combination of the $\overline{q}_{3}$ and $\overline{q}_{12}$ parameters provides the minimal mislabeling value and the largest separation between the clouds associated with hydrate- and liquid-like water molecules in the  $\overline{q}_{12}$ -- $\overline{q}_{3}$ parameter plane. We have found that this combination of local order parameters is able to distinguish between water molecules in hydrate and liquid phases of hydrates that exhibit sI and sII crystallographic structures at different thermodynamic conditions, including thermodynamic states at coexistence conditions between hydrate and liquid phases and at supercooling conditions.

Finally, we have investigated the effect of calculating averaged local order parameters from simulations that exhibit interfaces separating phases involving hydrate and aqueous solutions. In particular, we have determined the dependency of the $\overline{q}_{12}$ on the position of water molecules along simulation boxes containing interfaces hydrate--aqueous solution and aqueous solution--CO$_{2}$. We have also considered its dependence on the number of water neighbors. Our results indicate that the presence of interfaces in the simulation box significantly alters the local environment of water molecules and the criterion used to identify water molecules as solid-like or liquid-like fails. Particularly, order parameters take unrealistic high values for water molecules not surrounded by other water molecules, as it happens at interfaces in which one of the phases is not aqueous-like.  This produces a wrong identification of water molecules as hydrate-like. According to our results, a good receipt to solve the problem of non-physical high values of the order parameters in simulations of systems that exhibit this kind of interfaces is to identify water molecules using the threshold criterion of order parameters in combination with the number of neighbors a water molecule has. This method enables the identification of water molecules as hydrate-like without overestimating their number caused by non-physical values of the order parameters.

\section*{Conflicts of interest}

The authors have no conflicts to disclose.

\section*{Acknowledgments}
This work was funded by Ministerio de Ciencia e Innovaci\'on (Grant No.~PID2021-125081NB-I00) and Universidad de Huelva (P.O. FEDER EPIT1282023), both co-financed by EU FEDER funds. We greatly acknowledge RES resources provided by Barcelona Supercomputing Center in Mare Nostrum to FI-2023-2-0041 and FI-2023-3-0011 and by The Supercomputing and Bioinnovation Center of the University of M\'alaga in Picasso to FI-2024-1-0017. We also thank helpful discussions with Carlos Vega, Eduardo Sanz, and Samuel Blazquez. One of us (F.J.B.) would also like to dedicate this paper to the memory of Prof.~Luis F.~Rull. I met Luis on my way when I was studying Physics at the University of Seville, the Alma Matter of Luis.
He was my teacher in the Statistical Physics Third-year course and I received classes from him and Prof.~Lourdes F. Vega, PhD student of Luis by that time and my future PhD advisor. I was fascinated by the Gibbs ensembles' deep meaning and the possibility of simulating microscopic systems using Monte Carlo and molecular dynamics methods. The simulation school Luis (my scientific grandfather) created is still alive in the current generation working on Statistical Physics in Spain. I.M.Z. also thanks to Prof.~J.~L.~Abascal for all the knowledge and computational abilities shared with him during his first postdoctoral stay at the Complutense University of Madrid, where Prof.~Abascal co-supervised the research conducted at that prestigious university, which led to one of I.M.Z.'s most cited papers to date. Beyond academic reasons, I.M.Z. appreciates the kindness, humanity, and friendly personality of Prof.~Abascal shown day by day.

\section*{Data availability}

The data that support the findings of this study are available within the article.

\section*{References}
\bibliography{masterbib}

\end{document}